%% file: main.tex
\documentclass[sigconf,nonacm]{acmart}

\usepackage{stfloats}
\usepackage{booktabs} 
\usepackage{fontawesome}
\usepackage{changepage}

\usepackage{array}
\newcommand{\PreserveBackslash}[1]{\let\temp=\\#1\let\\=\temp}
\newcolumntype{C}[1]{>{\PreserveBackslash\centering}p{#1}}
\newcolumntype{R}[1]{>{\PreserveBackslash\raggedleft}p{#1}}
\newcolumntype{L}[1]{>{\PreserveBackslash\raggedright}p{#1}}

\usepackage[most]{tcolorbox}
\definecolor{niceRed}{RGB}{65, 65, 65}
\definecolor{niceBlue}{RGB}{82, 82, 179}

\usepackage{amsmath}
\usepackage{bm}
\usepackage{amsfonts}
\usepackage{amsthm}
\usepackage{multirow}
\usepackage{colortbl}
\usepackage{enumitem}
\usepackage{subcaption}
\usepackage[normalem]{ulem}
\usepackage{ragged2e}
\usepackage{pifont}

\usepackage{tikz}

\newcommand{\cmark}{\ding{51}}%
\newcommand{\xmark}{\ding{55}}%

\newtcbox{\pill}{on line,
  colframe={niceRed!30!white},
  colback=niceRed!5!white,
  boxrule=1pt,
  arc=3pt,boxsep=0pt,left=2pt,right=2pt,top=2pt,bottom=2pt}

\newtcbox{\pillMatch}{on line,
    sharp corners, 
    rounded corners=west,
  colframe=niceBlue,colback=niceBlue!5!white,
  boxrule=1pt,arc=3pt,boxsep=0pt,left=2pt,right=2pt,top=2pt,bottom=2pt,
  }
\newtcbox{\pillMatchScore}{on line,
    sharp corners, 
    rounded corners=east,
  colframe=niceBlue,colback=niceBlue,fontupper=\color{white},
  grow to left by=0.01cm,
  boxrule=1pt,arc=3pt,boxsep=0pt,left=2pt,right=2pt,top=3.6pt,bottom=2.45pt,
  }

\input{dlbook_math}

\usepackage{xspace}
\newcommand{\sref}[1]{\S\ref{#1}\xspace}
\usepackage{accents}
\newlength{\dhatheight}

\newcommand{\doublehat}[1]{%
    \settoheight{\dhatheight}{\ensuremath{\hat{#1}}}%
    \addtolength{\dhatheight}{-0.35ex}%
    \hat{\vphantom{\rule{1pt}{\dhatheight}}%
    \smash{\hat{#1}}}}

\DeclareMathOperator*{\bertdot}{BERT_\text{DOT}}
\DeclareMathOperator*{\bert}{BERT}

\DeclareMathOperator*{\colbert}{ColBERT}

\begin{document}

\title{Introducing Neural Bag of Whole-Words with ColBERTer: Contextualized~Late~Interactions~using~Enhanced~Reduction}

\newcommand{\tsc}[1]{\textsuperscript{#1}} 
\author{Sebastian Hofst{\"a}tter\tsc{1}, Omar Khattab\tsc{2}, Sophia Althammer\tsc{1}, Mete Sertkan\tsc{1}, Allan Hanbury\tsc{1}}
\affiliation{
  \institution{\vskip .15cm}
  \institution{\tsc{1} TU Wien, \tsc{2} Stanford University}
  \institution{\vskip .15cm}
}

\renewcommand{\authors}{Sebastian Hofst{\"a}tter, Omar Khattab, Sophia Althammer, Mete Sertkan, Allan Hanbury}

\begin{abstract}

Recent progress in neural information retrieval has demonstrated large gains in effectiveness, while often sacrificing the efficiency and interpretability of the neural model compared to classical approaches. This paper proposes ColBERTer, a neural retrieval model using contextualized late interaction (ColBERT) with enhanced reduction. Along the effectiveness Pareto frontier, ColBERTer's reductions dramatically lower ColBERT's storage requirements while simultaneously improving the interpretability of its token-matching scores. To this end, ColBERTer fuses single-vector retrieval, multi-vector refinement, and optional lexical matching components into one model. For its multi-vector component, ColBERTer reduces the number of stored vectors per document by learning unique whole-word representations for the terms in each document and learning to identify and remove word representations that are not essential to effective scoring. We employ an explicit multi-task, multi-stage training to facilitate using very small vector dimensions.
Results on the MS MARCO and TREC-DL collection show that ColBERTer can reduce the storage footprint by up to 2.5$\times$, while maintaining effectiveness. With just one dimension per token in its smallest setting, ColBERTer achieves index storage parity with the plaintext size, with very strong effectiveness results. Finally, we demonstrate ColBERTer's robustness on seven high-quality out-of-domain collections, yielding statistically significant gains over traditional retrieval baselines.

\end{abstract}

\maketitle

\input{sections/1.intro}
\input{sections/2.background}
\input{sections/3.colberter}

\input{sections/4.lifecycle}
\input{sections/5.experiment.design}
\input{sections/6.results}

\input{sections/7.conclusion}


\bibliographystyle{ACM-Reference-Format}
\bibliography{my-references}


\end{document}

%% file: dlbook_math.tex
\def\eqref#1{equation~\ref{#1}}

\def\1{\bm{1}}

\DeclareMathAlphabet{\mathsfit}{\encodingdefault}{\sfdefault}{m}{sl}
\SetMathAlphabet{\mathsfit}{bold}{\encodingdefault}{\sfdefault}{bx}{n}



%% file: sections/1.intro.tex
\section{Introduction}
\label{sec:intro}

%
%
Traditional retrieval systems have long relied on bag-of-words representations to search within unstructured text collections. This has led to mature architectures that are known to support highly-efficient search. The compact inverted indexes enable fast top-$k$ retrieval strategies, while also exhibiting interpretable behavior, where retrieval scores can directly be attributed to the contributions of individual terms. Despite these attractive qualities, recent progress in Information Retrieval (IR) has firmly demonstrated that pre-trained language models can considerably boost retrieval quality over classical approaches. This progress has not come without downsides, as it is less clear how to control the computational cost and how to ensure interpretability of these neural models. This has sparked an unprecedented tension in IR between achieving the best retrieval quality, maintaining low computational costs, and prioritizing interpretable modeling.

\begin{figure}[t]
    
    \begin{FlushLeft}
    \faSearch \hspace{0.1cm} \textbf{does doxycycline contain sulfa}
    \end{FlushLeft}

    \begin{FlushLeft}
    \begin{adjustwidth}{0.5cm}{}
    \textit{BERT tokenized (9 subword-tokens):  'does', 'do', '\#\#xy',  \\ '\#\#cy',
    '\#\#cl', '\#\#ine', 'contain', 'sul', '\#\#fa'}
    \end{adjustwidth}
    \end{FlushLeft}
    \vspace{0.1cm}
    \vspace{0.1cm}
    \begin{FlushLeft}
    \textbf{ColBERTer BOW$^2$} \textit{(30 saved vectors from 84 subword-tokens)}:

    \vspace{0.1cm}
\pill{\vphantom{Hy}photosensitivity}
\pillMatch{\vphantom{Hy}doxycycline}\pillMatchScore{\footnotesize \vphantom{Hy}12.9}
\pillMatch{\vphantom{Hy}sulfa}\pillMatchScore{\footnotesize \vphantom{Hy}14.2}
\pill{\vphantom{Hy}sunburned}
\pill{\vphantom{Hy}rash}
\pill{\vphantom{Hy}clothing}
\pill{\vphantom{Hy}sunlight}
\pill{\vphantom{Hy}allergic}
\pill{\vphantom{Hy}compound}
\pill{\vphantom{Hy}drugs}
\pillMatch{\vphantom{Hy}containing}\pillMatchScore{\footnotesize \vphantom{Hy}6.6}
\pill{\vphantom{Hy}take}
\pill{\vphantom{Hy}safely}
\pill{\vphantom{Hy}wear}
\pill{\vphantom{Hy}.}
\pill{\vphantom{Hy}is}
\pillMatch{\vphantom{Hy}no}\pillMatchScore{\footnotesize \vphantom{Hy}4.7}
\pill{\vphantom{Hy}exposed}
...
    \vspace{-0.1cm}

    \small{\textbf{Fulltext:} No doxycycline is not a sulfa containing compound, so you may take it safely if you are allergic to sulfa drugs. You should be aware, however, that doxycycline may cause photosensitivity, so you should wear appropriate clothing, or you may get easily sunburned or develop a rash if you are exposed to sunlight.}
    \end{FlushLeft}
%
%
%
%
%
    \vspace{-0.2cm}
    \caption{Example of ColBERTer's BOW$^2$ (Bag Of Whole-Words): ColBERTer stores and matches unique whole-word representations. 
    The words in BOW$^2$ are ordered by implicitly learned query-independent term importance. Matched words are highlighted in blue with whole-word scores displayed in a user-friendly way next to them.}
    \label{fig:hero_example}
    \vspace{-0.4cm}
\end{figure}

%
%
For practical applications, ranking model architectures are confined to strict cost constraints, primarily query latency and index space footprint. While a larger disk space consumption might not be a critical cost factor, keeping many pre-computed representations in memory---as often needed for low query latency---does increase hardware costs significantly. For multi-vector models like ColBERT~\cite{khattab2020colbert}, space consumption is determined by a multiplication of three variables: 1) the number of vectors per document; 2) the number of dimensions per vector; 3) the number of bytes per dimension. A motivating observation of this work is that reducing any of these three variables by a certain factor directly reduces the storage requirement by that factor and yet this does not necessary translate to the same ratio of keeping or reducing the effectiveness. Well-studied \textit{low hanging fruits} for good tradeoffs include reducing the number of dimensions (either inside the model or outside with PCA) and reducing the number of bytes with quantization~\cite{ma2021simple,ji2019efficient,lewis2021boosted,gao2021coil}. The remaining parameter, the number of vectors per document, offers many development opportunities determined by the model architecture and the retrieval approach employed. 

%
%
In addition to efficiency considerations, with the accelerating adoption and impact of machine learning models, there are indications that future regulatory environments will require deployed models to provide transparent and reliably interpretable output to their users.\footnote{Such as a recent 2021 proposal by the EU Commission on AI regulation, see: \\ \url{https://eur-lex.europa.eu/legal-content/EN/TXT/?uri=CELEX:52021PC0206} (Art. 13)} This crucial need for interpretability is especially pronounced in IR, where the ranking models are demanded to be fair and transparent \cite{castello2019fairranking}. Despite this, we observe that the two largest classes of neural models at the moment---namely, cross-encoders and single-vector bi-encoders---rely on opaque aggregation functions which do not expose transparently the contributions of the terms of a query or passage to the score.

%
%
This paper presents a novel end-to-end retrieval and ranking model called \textbf{ColBERTer}. ColBERTer extends the popular ColBERT model with effective \underline{\textbf{e}}nhanced \underline{\textbf{r}}eduction approaches. These enhanced reductions increase the level of interpretability and reduce the storage and latency cost greatly, while at the same time maintaining the quality of retrieval. 

%
%

Our first main architectural contribution is fusing a single-vector retrieval and multi-vector refinement model into one with explicit multi-task training. \citet{mackenzie2021wacky} showed that fitting subword models into established retrieval systems negatively impacts the performance of previously developed efficient query evaluation techniques, such as early exiting.
Therefore, we introduce neural Bag of Whole-Words (BOW$^2$) representations for increasing interpretability and reducing the number of stored vectors in the ranking process. The BOW$^2$ consist of the aggregation of all subword token representations contained in a unique whole word. 

In order to further reduce the number of stored vectors we learn to reduce the BOW$^2$ representations with simplified contextualized stopwords (CS) \cite{Hofstaetter2020_cikm}.
To maximally reduce the dimensionality of the token vectors to one, we also employ an Exact Matching (EM) component, which matches only the vector representations of whole-words which are exact matches from the query. We refer to this model variant as \textbf{Uni-ColBERTer} (following the nomenclature of \citet{lin2021few}).

%
%
In Figure \ref{fig:hero_example} we show ColBERTer's BOW$^2$ representation and how we can display whole-word scores to the user in a keyword view. As we are aggregating all subwords to whole-words, we can confidently display the whole-word scores of this complex medical-domain query to show ColBERTer's interpretability capabilities, without cherry picking an example that only contains words that are fully part of BERT's vocabulary.

The ColBERTer encoding architecture enables many different indexing and retrieval scenarios. Building on recent works \cite{gao2021coil,lin2021few}, which already proposed and analyzed some dense and sparse retrieval modes we provide a holistic categorization and ablation study of five possible usage scenarios of ColBERTer encoded sequences:  sparse token retrieval, dense single vector retrieval, as well as refining either one of the retrieval sources and a full hybrid mode.
%
%
Specifically, we study for our ColBERTer model:

\newcommand{\RQone}{\begin{itemize}
    \item[\textbf{RQ1}] Which aggregation and training regime works best for combined retrieval and refinement capabilities of ColBERTer?
\end{itemize}}
\RQone

We find that multi-task learning with two weighted loss functions for the retrieval and refinement and a learned score aggregation of the retrieval and refinement score outperforms consistently fixed score aggregation. Furthermore we investigate the joint training of score aggregation, BOW$^2$, and contextualized stopwords with a weighted multi-task loss. We find that tuning the different weights does improve the tradeoff between removed vectors and retrieval quality, however overall the results are robust to small changes in the hyperparameter space.

Following our holistic definition of dense and sparse combinations and to provide guidance on how to employ ColBERTer we study various deployment scenarios and answer:  
\newcommand{\RQthree}{\begin{itemize}
    \item[\textbf{RQ2}] What is the best indexing and refinement strategy for ColBERTer?
\end{itemize}}
\RQthree

Interestingly, we find that a full hybrid retrieval deployment is not practically necessary, and only results in very modest and not significant gains compared to either a sparse or dense index with passage refinement of the other component. Using a dense index results in higher recall than a sparse index, however after the initial retrieval results are refined, the effect on the top $10$ results becomes negligible -- especially on TREC-DL. 

This novel result could lead to less complexity in deployment, as only one index is required. Practitioners could choose to keep a sparse index, if they already made significant investments or choose only a dense approximate nearest neighbor index for more predictable query latency. Both sparse and dense encodings of ColBERTer can be optimized with common indexing improvements.  

With our hyperparameters fixed, we aim to understand the quality effect of reducing storage factors along 2 axes of ColBERTer:

\newcommand{\RQtwo}{\begin{itemize}
    \item[\textbf{RQ3}] How do different configurations of dimensionality and vector count affect the retrieval quality of ColBERTer?
\end{itemize}}
\RQtwo

We study the effect of BOW$^2$, CS, and EM reductions on different token vector dimensionalities ($32$, $16$, $8$, and $1$). We find that, as expected, retrieval quality is reduced with each dimension reduction, however the delta is small. Furthermore, we observe that BOW$^2$ and CS reductions result -- on every dimension setting -- in a Pareto improvement over simply reducing the number of dimensions.

While we want to emphasize that it becomes increasingly hard to contrast neural retrieval architectures -- due to the diversity surrounding training procedures -- and make conclusive statements about "SOTA" -- due to evaluation uncertainty -- we still compare ColBERTer to related approaches:
\newcommand{\RQfive}{\begin{itemize}
    \item[\textbf{RQ4}] How does the fully optimized ColBERTer system compare to other end-to-end retrieval approaches?
\end{itemize}}
\RQfive

We find that ColBERTer does improve effectiveness compared to related approaches, especially for cases with low storage footprint. Our Uni-ColBERTer variant especially outperforms previous single-dimension token encoding approaches, while at the same time offering improved transparency and making it easier to showcase model scores with mappings to whole words. 

In order to evaluate the generalizability of ColBERTer we study its results on seven high-quality and diverse retrieval collections from different domains using a meta-analysis \cite{soboroff2018meta}, that allows us to robustly observe whether statistical significant gains are achieved over multiple collections or not. We investigate:

\newcommand{\RQsix}{\begin{itemize}
    \item[\textbf{RQ5}] How robust is ColBERTer when applied to out of domain collections?
\end{itemize}}
\RQsix

We find that ColBERTer with token embeddings of $32$ or Uni-ColBERTer with $1$ dimension both show an overall significantly higher retrieval effectiveness compared to BM25, with not a single collection worse than BM25. Compared to a TAS-Balanced trained dense retriever \cite{Hofstaetter2021_tasb_dense_retrieval} ColBERTer is not statistically significantly worse on any single collection. While we observe an overall positive effect it is not statistically significant within a $95\%$ confidence interval. This robust analysis tries to not overestimate the benefits of ColBERTer, while at the same time giving us more confidence in the results. 

We publish our code, trained models, and documentation at: \\ \textit{github.com/sebastian-hofstaetter/colberter}

%% file: sections/2.background.tex
\section{Background}

In this section we motivate storing unique whole-word representations instead of all sub-word tokens with statistics of our collections; we review the single-vector $\bertdot$ and multi-vector $\colbert$ architectures; and give an overview of other related approaches.

\subsection{Tokenization}

We use, as most other neural ranking approaches, a BERT \cite{devlin2018bert} variant to contextualize sequences. Therefore, we are locked into the specific tokenization the language model uses. The BERT tokenization process first splits full text on all whitespace and punctuation characters, leaving whole words intact. In the second step it uses the WordPiece algorithm \cite{schuster2012japanese} to split words to sub-word tokens in the reduced vocabulary. In Table \ref{tab:word_stats} we showcase word count statistics based on the final sub-word tokens as well as the whole-words from the first step. We count all units as well as unique sets of lowercased and stemmed words. We can clearly observe that aggregating unique+stemmed whole-words only save from 59 \% to 36 \% of the original sub-word units that correspond to the BERT output. Related multi-vector methods, such as the ColBERT or (Uni)COIL models save all BERT tokens (leftmost column), while the BOW$^2$ aggregation we propose in \sref{sec:bow2} saves only stemmed unique whole-words (rightmost column).

\begin{table}[t]
    \centering
    \caption{Average token count statistics per passage for our used collections using different tokenization and aggregation approaches. \textit{\% Ret. refers to the percent of retained vectors between all BERT tokens and unique+stemmed whole words.} }
    \label{tab:word_stats}
    \setlength\tabcolsep{2.0pt}
    \vspace{-0.3cm}
    \begin{tabular}{l!{\color{gray}\vrule}rr!{\color{lightgray}\vrule}rrr!{\color{lightgray}\vrule}r}
       
       \toprule

       \multirow{2}{*}{\textbf{Collection}} &
       \multicolumn{2}{c!{\color{lightgray}\vrule}}{\textbf{BERT Tokens}}&
       \multicolumn{3}{c!{\color{lightgray}\vrule}}{\textbf{Whole-Words}}&\textbf{\%}\\

       & All & Unique & All & Unique & U.+Stem & \textbf{Ret.} \\
       \midrule
       \arrayrulecolor{lightgray}
        \textbf{MS MARCO} & 76.9 & 50.2 & 68.6 & 44.1 & 43.2 & \textbf{56 \%}\\

        \midrule
        
        \textbf{TREC Covid} & 237.7 & 119.5 & 199.5 & 98.7 & 94.3       & \textbf{40 \%}\\
        \textbf{TripClick} & 380.2 & 173.1 & 324.1 & 144.6 & 137.6      & \textbf{36 \%}\\
        \textbf{NFCorpus} & 348.0 & 164.4 & 296.6 & 137.4 & 131.2       & \textbf{37 \%}\\
        \textbf{DBPedia Entity} & 72.8 & 49.1 & 61.0 & 41.0 & 40.5      & \textbf{55 \%}\\
        \textbf{Antique} & 53.1 & 36.4 & 48.3 & 32.1 & 31.5             & \textbf{59 \%}\\
        \textbf{TREC Podcast} & 421.2 & 183.1 & 409.4 & 173.2 & 166.6   & \textbf{40 \%}\\
        \textbf{TREC Robust 04} & 379.9 & 191.9 & 363.5 & 172.1 & 165.0 & \textbf{43 \%}\\

        \arrayrulecolor{black}
       \bottomrule
    \end{tabular}
\end{table}

\subsection{BERT\texorpdfstring{$_\textbf{DOT} $}: Architecture}

The BERT$_\text{DOT}$ model matches a single vector of the query with a single vector of a passage. The vectors are the output of independent $\bert$ computations as follows:
\begin{equation}
\begin{aligned} 
\tilde{q} &= \bert([\text{CLS};{q}_{1:m};\text{SEP}])_\text{CLS} * W_d  \\
\tilde{p} &= \bert([\text{CLS};{p}_{1:n};\text{SEP}])_\text{CLS} * W_d
\end{aligned}
\end{equation}
where we pool a single vector as the index selection corresponding to the CLS vector and optionally compress the vector dimensionality with a single shared linear layer $W_d$. This setup allows us to pre-compute every contextualized passage representation $\tilde{p}$. Then, either the model or a nearest neighbor index, computes the final scores as the dot product $\cdot$ of $\tilde{q}$ and $\tilde{p}$:
\begin{equation}
\begin{aligned}
    \bertdot({q}_{1:m},{p}_{1:n}) & = \tilde{q} \cdot \tilde{p}
\end{aligned}
\end{equation}

While the model architecture itself does not allow for many modifications, the training regime is a widely studied way of improving the results of the single vector retrieval platform \cite{xiong2020approximate,luan2020sparse,lu2020twinbert}.

\subsection{ColBERT Architecture}
\label{sec:colbert}

%
%
The $\colbert$ model \cite{khattab2020colbert} delays the interactions between query and document to after the BERT computation. $\colbert$ contextualizes every query and document subword representation, by feeding the subword-tokenized query ${q}_{1:m}$ and passage ${p}_{1:n}$ sequences through a BERT model:
\begin{equation}
\begin{aligned} 
\tilde{q}_{1:m+r+2} &= \bert([\text{CLS};{q}_{1:m};\operatorname{rep}(\text{MASK},r);\text{SEP}]) * W_d  \\
\tilde{p}_{1:n+2} &= \bert([\text{CLS};{p}_{1:n};\text{SEP}]) * W_d
\label{eq:colbert_1}
\end{aligned}
\end{equation}
where the $\operatorname{rep}(MASK,r)$ method repeats the MASK token a number of times, set by the hyperparameter $r$. \citet{khattab2020colbert} introduced this query augmentation method to increase the computational capacity of the BERT model for short queries. The dimensions of all representation vectors are reduced by a single shared linear layer $W_d$.
The interactions in the $\colbert$ model are aggregated with a max-pooling per query term and sum of query-term scores as follows:
\begin{equation}
\begin{aligned}
    \colbert({q}_{1:m},{p}_{1:n}) = \sum_{1}^{m+r+2} \max_{1..n+2} \tilde{q}_{1:m+r+2}^T \cdot \tilde{p}_{1:n+2}
\label{eq:colbert_2}
\end{aligned}
\end{equation}
The max-sum operator scans the matrix of all term-by-term interactions, which is a technique inspired by earlier works on kernel-pooling \cite{Xiong2017,Hofstaetter2020_ecai}. The term-by-term interaction matrix creates transparency in the scoring, as it allows to inspect the source of different scoring parts, while being mapped to human-readable word units \cite{Hofstaetter2020_ecir}. However, the usefulness of this feature is reduced by the use of special tokens, especially by the query expansion with MASK tokens, as it is non-trivial to explain reliably to the users what each MASK token stands for.

\begin{figure*}[t]
   \includegraphics[clip,trim={0.2cm 0.2cm 0.2cm 0.2cm}, width=0.98\textwidth]{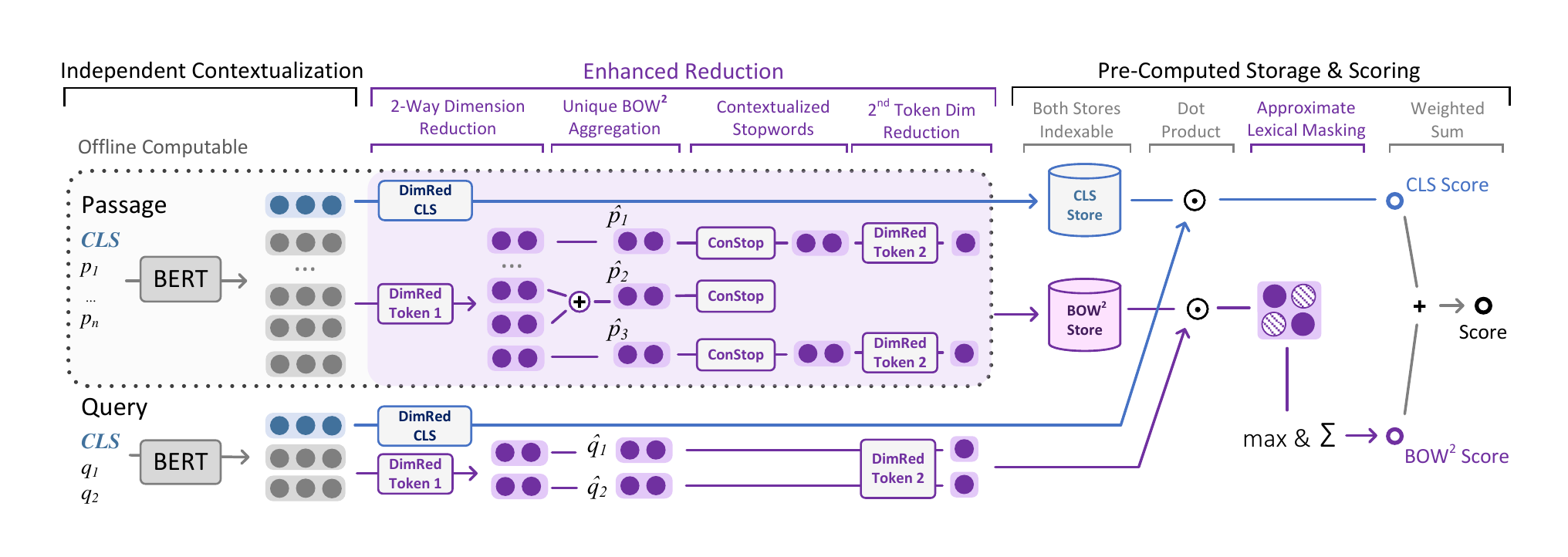}
    \centering
    \vspace{-0.3cm}
    \caption{The ColBERTer encoding architecture, followed by the query-time workflow. The passage representations (both the single CLS and token vectors) are pre-computed during indexing time. The enhanced reductions with the 2-way dimension reduction, the unique BOW$^2$ aggregation, contextualized stopwords and token dimensionality reduction are applied symmetrically to passages and queries (except for the stopword removal).}
    \label{fig:colberter_architecture}
    \vspace{-0.1cm}
\end{figure*}

\subsection{Related Work}

\paragraph{\textbf{Vector Reduction}}

Our dynamic approach to reduce the number of vectors needed to represent a passage differs from previous works that focus on fixed numbers of vectors across all passages: \citet{lassance2021study} prune ColBERT representations to either 50 or 10 vectors for MSMARCO by sorting tokens either by Inverse Document Frequency (IDF) or the last-layer attention scores of BERT. \citet{zhou2021multi} extend ColBERT with temporal pooling, by sliding a window over the passage representations to create a representation vector every window size steps, with a fixed target count of representation vectors.
\citet{luan-etal-2021-sparsemebert} represent each passage with a fixed number of contextualized embeddings of the CLS token and the first $m$ token of the passage and score the relevance of the passage with the maximum score of the embeddings. \citet{humeau2020polyencoders} compute a fixed number of vectors per query, and aggregate them by softmax attention against document vectors.
\citet{lee2020learning} learn phrase (multi-word) representations for QA collections. This approach also reduces the vector count compared to indexing all tokens, however it fully depends on the availability of exact answer spans in the passages and is therefore not universally applicable in the IR setting. 
\citet{tonelloto2021embeddingpruning} prune the embeddings of the query terms, however they do not reduce the number of embeddings for the documents in the index.

In summary, the clearest difference from our contributions to related vector reduction techniques is: 1) we reduce a dynamic number of vectors per passage; 2) we keep a mapping between human-readable tokens and vectors, allowing scoring information to be used in the user interface; 3) we learn the full pruning process end-to-end without term-based supervision. 

\paragraph{\textbf{Vector Compression}}

\citet{ma2021simple} study various learning and post-processing methods to reduce dimensionality of dense retrieval vectors. In contrast to our study, they find that learned dimensionality reduction performs poorly. Also for single vector retrieval \citet{zhan2021jointly} optimize product quantization as part of the training.
Recently, \citet{santhanam2021colbertv2} study residual compression of all saved token vectors as part of the ColBERT end-to-end retrieval setting. 
There are concurrent efforts revisiting lexical matching with learned sparse representations  \cite{gao2021coil,formal2021splade,lin2021few}, which employ the efficiency of exact lexical matches. Different to our work, they focus on reducing the number of dimensions of the learned embeddings without reducing the number of stored tokens. Many of these approaches can be considered complementary to our proposed methods, and future work should evaluate how well these methods compose to achieve even larger compression rates.

%% file: sections/3.colberter.tex
\section{\texorpdfstring{C\lowercase{ol}BERT\lowercase{er}: Enhanced Reduction}.}

ColBERT with enhanced reduction, or \textbf{ColBERTer}, combines the encoding architectures of $\bertdot$ and $\colbert$, while extremely reducing the token storage and latency requirements along the effectiveness Pareto frontier. Our enhancements maintain model transparency, creating a concrete mapping of scoring sources and human-readable whole-words. We provide an overview of the encoding and retrieval workflow in Figure~\ref{fig:colberter_architecture}.

ColBERTer independently encodes the query and the document using a transformer encoder like BERT, producing token-level representation similar to ColBERT:
\begin{equation}
\begin{aligned} 
\tilde{q}_{1:m+2} &= \bert([\text{CLS};{q}_{1:m};\text{SEP}]) \\
\tilde{p}_{1:n+2} &= \bert([\text{CLS};{p}_{1:n};\text{SEP}])
\label{eq:colberter_1}
\end{aligned}
\end{equation}
To maximize transparency, we do not apply the query augmentation mechanism of \citet{khattab2020colbert} (see~\sref{sec:colbert}), which appends MASK tokens to the query with the goal of implicit -- and thus potentially opaque -- query expansion.

\subsection{2-Way Dimension Reduction}
\label{sec:dim-red}

Given the transformer encoder output, ColBERTer uses linear layers to reduce the dimensionality of the output vectors in two ways: 1) we use the linear layer $W_{CLS}$ to control the dimension of the first \texttt{CLS}-token representation (e.g. $128$ dimensions): 
\begin{equation}
\begin{aligned} 
{q}_{CLS} &= \tilde{q}_{1} * W_{CLS}  \\
{p}_{CLS} &= \tilde{p}_{1} * W_{CLS}
\label{eq:colberter_dim_red_cls}
\end{aligned}
\end{equation}
and 2) the layer $W_t$ projects the representations of the remaining tokens down to the token embedding dimension (usually smaller, e.g. $32$):
\begin{equation}
\begin{aligned} 
\dot{q}_{1:m} &= \tilde{q}_{2:m+1} * W_t  \\
\dot{p}_{1:n} &= \tilde{p}_{2:n+1} * W_t
\label{eq:colberter_dim_red_token}
\end{aligned}
\end{equation}

This 2-way reduction combined with our novel training workflow (\sref{sec:training_proc}) serves to reduce our space footprint compared to ColBERT and at the same time provides more expressive encodings than a single vector $\bertdot$ model. Furthermore, it enables a multitude of potential dense and sparse retrieval workflows (\sref{sec:retrieval_workflows}).

\subsection{\texorpdfstring{BOW$^\textbf{2}$} :: Bag of Unique Whole-Words}
\label{sec:bow2}

Given the token representations ($\dot{q}_{1:m}$ and $\dot{p}_{1:n}$), ColBERTer applies its novel key transformation: BOW$^2$ to the sequence of vectors. Whereas ColBERT and COIL maintain one vector for each BERT token, including tokens corresponding to sub-words in the BERT vocabulary, we create a single representation for each unique whole word. This serves to further reduce the storage overhead of our model by reducing the number of tokens, while preserving an explicit mapping of score parts to human understandable words.

During tokenization we build a mapping between each sub-word token and corresponding unique whole word (as defined by a simple split on punctuation and whitespace characters). The words can also be transformed through classical IR techniques such as stemming. Then, inside the model we aggregate whole word representations for each whole word $w$ in passage $p$ by computing the mean of the embeddings of $w$'s constituent sub-words $\dot{p}_i$. We get the set of unique whole-word representation of the passage $p$:

\begin{equation}
\begin{aligned} 
\hat{p}_{1:\hat{n}} &= \bigg\{ \frac{1}{|\dot{p}_i \in w|} \sum_{\dot{p}_i \in w} \dot{p}_i \ \bigg| \  \forall \ w \in \operatorname{BOW^2}(p) \bigg\}
\label{eq:colberter_bow_aggreagtion}
\end{aligned}
\end{equation}

We apply the same procedure symmetrically to the query vectors $\dot{q}_{1:m}$ from equation (7) as well to produce $\hat{q}_{1:\hat{m}}$.
The resulting sets are still dynamic in length as their length now depends on the number of whole words ($\hat{n}$ and $\hat{m}$ for passage and query sequences respectively). We refer to the new sets as \textit{bag of words}, as we only save one word once and the order of the vectors now does not matter anymore, because the language model contextualization already happened.

\subsection{Simplified Contextualized Stopwords}

To further reduce the number of passage tokens to store, we adopt a simplified version of \citet{Hofstaetter2020_cikm}'s contextualized stopwords (CS), which was first introduced for the TK-Sparse model. CS learns a \textit{removal gate} of tokens solely based on their context-dependent vector representations. We simplify the original implementation of CS and adapt the removal process to fit into the encoding phase of the ColBERTer model. 

Every whole-word passage vector $\hat{p}_j$ is transformed by a linear layer (with weight matrix $W_{s}$ and bias vector $b_{s}$), followed by a ReLU activation, to compute a single-dimensional stopword removal gate $r_j$:
\begin{equation}
r_{j} = \operatorname{ReLU}(\hat{p}_{j} W_s + b_s)
\label{eq:stopword_gate}
\end{equation}

The original implementation \cite{Hofstaetter2020_cikm} masks scores after TK's kernel-activation, meaning the non-zero gates have to be saved as well, which increases the systems' complexity. In contrast, we directly apply the gate to the representation vectors. In particular, we drop every representation where the gate $r_j=0$, and otherwise scale the magnitude of the remaining representations using their gate scores:
\begin{equation}
\hat{p}_j = \hat{p}_j * \hat{r}_{j}
\label{eq:stopword_gate_application}
\end{equation}

This fully differentiable approach allows us to learn the stopword gate during training and remove all nullified vectors at indexing time, as they do not contribute to document scores. Applying the stopword gate directly to the representation vector allows us to observe much more stable training than the authors of TK-Sparse observed -- we do not need to adapt the training procedure with special mechanisms to keep the model from collapsing. Following \citet{Hofstaetter2020_cikm} we train the removal gate with a regularization loss, which forces the stopword removal gate to become active during training as described in \sref{sec:training_proc}.

\subsection{Matching \& Score Aggregation}
\label{sec:matching}

After we complete the independent encoding of query and passage sequences, we need to match and score them. ColBERTer creates two scores, one for the CLS vector and one for the token vectors.
The CLS score is a dot product of the two CLS vectors:
\begin{equation}
\begin{aligned}
    s_{CLS} = {q}_{CLS} \cdot {p}_{CLS}
\label{eq:colberter_cls_match}
\end{aligned}
\end{equation}
The token score follows the scoring regime of ColBERT, with a match matrix of word-by-word dot product and max-pooling the document word dimension followed by a sum over all query words:

\begin{equation}
\begin{aligned}
    s_{token} = \sum_{j=1}^{\hat{m}} \max_{i=1..\hat{n}} \hat{q}_{j}^T \cdot \hat{p}_{i}
\label{eq:colberter_token_match}
\end{aligned}
\end{equation}

The final score of a query-passage pair is computed with a learned aggregation of the two score components:
\begin{equation}
\begin{aligned}
    s_{ColBERTer} = \sigma(\gamma) * s_{CLS} + (1-\sigma(\gamma)) * s_{token}
\label{eq:learned_score_agg}
\end{aligned}
\end{equation}
where $\sigma$ is the sigmoid function, and $\gamma$ is a trainable scalar parameter. For the purpose of ablation studies $\sigma(\gamma)$ can be set to a fixed number, such as $0.5$. At first glance the learned weighting factor seems superfluous, as the upstream linear layers could already learn to change the magnitudes of the two components. However, we show in \sref{sec:source_of_effect} that the explicit weighting is crucial for the correct functioning of both components.

\subsection{\textbf{Uni-ColBERTer}: Extreme Reduction with Lexical Matching} 

While ColBERTer already is able to considerably reduce the dimensionality of the token representations, we found in pilot studies that for an embedding dimension of $8$ or lower the full match matrix is detrimental to the effectiveness. \citet{lin2021few} showed that a token score model can be effectively reduced to one dimension in UniCOIL. This effectively reduces the token representations to scalar \textit{weights}, necessitating an alternative mechanism to match query tokens with ``similar'' document tokens.

To fit the same reduction we need to apply more techniques to our ColBERTer architecture to create Uni-ColBERTer with single dimensional whole word vectors. While we now occupy the same bytes per vector, our vector reduction techniques make Uni-COLBERTer 2.5 times smaller than UniCOIL (on MSMARCO).

To reduce the token encoding to 1 dimension we apply a second linear layer after the contextualized stopword component:
\begin{equation} 
\begin{aligned} 
\doublehat{q}_{1:m+2} &= \hat{q}_{1:\hat{m}} * W_u  \\
\doublehat{p}_{1:n+2} &= \hat{p}_{1:\hat{n}} * W_u
\label{eq:colberter_dim_red_token_two}
\end{aligned}
\end{equation}

%
%

Furthermore, we need to apply a lexical match bias, following COIL's approach to only match identical words with each other.
This however brings an interesting engineering challenge: We do not build a global vocabulary with ids of whole-words during training nor inference. This is because doing so would be complex: Modern GPUs are so fast that to be able to saturate them we need multiple CPU processes (4-10 depending on the system) that prepare the input with tokenization, data transformation, and subsequent tensor batching of sequences. To keep track of a global vocabulary, these CPU processes would need to synchronize with a read-write dictionary on every token. This is simply not feasible in python multiprocessing while keeping the necessary speed to fully use even a single GPU. 

To overcome this problem, while still inducing an exact match bias, we propose approximate lexical interactions, by creating an n-bit hash $H$ from every whole-word without accounting for potential collisions and applying a mask of equal hashes to the match matrix. Depending on the selection of bits to keep this introduces different numbers of collisions.\footnote{In the case of MSMARCO we found the first 32 bits of sha256 to produce very few collisions (303 collisions out of 1.6 million hashes).} Depending on the collection size one can adjust the number of bits to save from the hash.
With the hashed global id of whole words we can adjust the match matrix of whole-words for low dimension token models as follows:

\begin{equation}
\begin{aligned}
    s_{token} = \sum_{1}^{\hat{m}} \max_{1..\hat{n} | H(w_{\hat{n}}) = H(w_{\hat{m}})} \doublehat{q}_{1:\hat{m}+2}^T \cdot \doublehat{p}_{1:\hat{n}+2}
\label{eq:colberter_approx_lexical_match}
\end{aligned}
\end{equation}

In practice, we implement this procedure by masking the full match matrix, so that the operation works on batched tensors. Besides allowing us reduce the token dimensionality to one, the lexical matching component of Uni-ColBERTer enables the sparse indexing of tokens in an inverted index, following UniCOIL.

%% file: sections/4.lifecycle.tex
\section{Model Lifecycle}

In this section we describe how we train our ColBERTer architecture and how we can deploy the trained model into a retrieval system.

\subsection{Training Workflow}
\label{sec:training_proc}

We train our ColBERTer model with triples of one query, and two passages where one is more relevant than the other. To incorporate the degree of relevance, as provided by a teacher model we use the Margin-MSE loss \cite{hofstaetter2020_crossarchitecture_kd}, formalized as follows:
\begin{equation}
\begin{aligned} 
\mathcal{L}_{MarginMSE}(M_s) = \operatorname{MSE}(&M_s^{+} - M_s^{-},\\ &M_t^{+} - M_t^{-}))
\end{aligned}
\end{equation}
Where a teacher model $M_t$ provides a teacher signal for our student model $M_s$ (in our case ColBERTer's output parts).

From the outside ColBERTer looks and acts like a single model, however it is in essence a multi-task model: aggregating sequences into a single vector, representing individual words, and actively removing uninformative words. Therefore, we need to train these three components in a balanced form, with a combined loss function as follows:
\begin{equation}
\mathcal{L} = \alpha_b * \mathcal{L}_b + \alpha_{CLS} * \mathcal{L}_{CLS} + \alpha_{CS} * \mathcal{L}_{CS}
\end{equation}
where $\alpha$'s are hyperparamters governing the weighting of the individual losses, which we explain in the following. 

The combined loss for both sub-scores $\mathcal{L}_b$ uses MarginMSE supervision on the final score:
\begin{equation}
\mathcal{L}_b = \mathcal{L}_{MarginMSE}(s_{ColBERTer})
\end{equation}

In pilot studies and shown in \sref{sec:source_of_effect} we observed that training ColBERTer only with a combined loss strongly reduces the effectiveness of the CLS vector alone. To overcome this issue and be able to use single vector retrieval we define $\mathcal{L}_{CLS}$ as:
\begin{equation}
\mathcal{L}_{CLS} = \mathcal{L}_{MarginMSE}(s_{CLS})
\end{equation}

Finally, to actually force the model to learn sparsity in the removal gate vector $r$ of the contextualized stopword component, we follow \citet{Hofstaetter2020_cikm} and add an $\mathcal{L}_{CS}$ loss of the L1-norm of the positive \& negative $r$:
\begin{equation}
\mathcal{L}_{CS} = ||r^{+}||_1 + ||r^{-}||_1
\end{equation}
By minimizing this loss, we introduce tension into the training process, as the sparsity loss needs to move as many entries to zero or close to zero, while the token loss as part of $\mathcal{L}_b$ needs non-zero entries to be able to determine relevance matches. To reduce volatility during training we train the enhanced reduction components one after another: We start with a ColBERT checkpoint, followed by the 2-way dimensionality reduction, BOW$^2$ and CS, and finally for Uni-ColBERTer we apply another round of dimensionality reduction. 

\begin{figure}[t]
   \includegraphics[clip,trim={0.2cm 0.2cm 0.2cm 0.7cm}, width=0.45\textwidth]{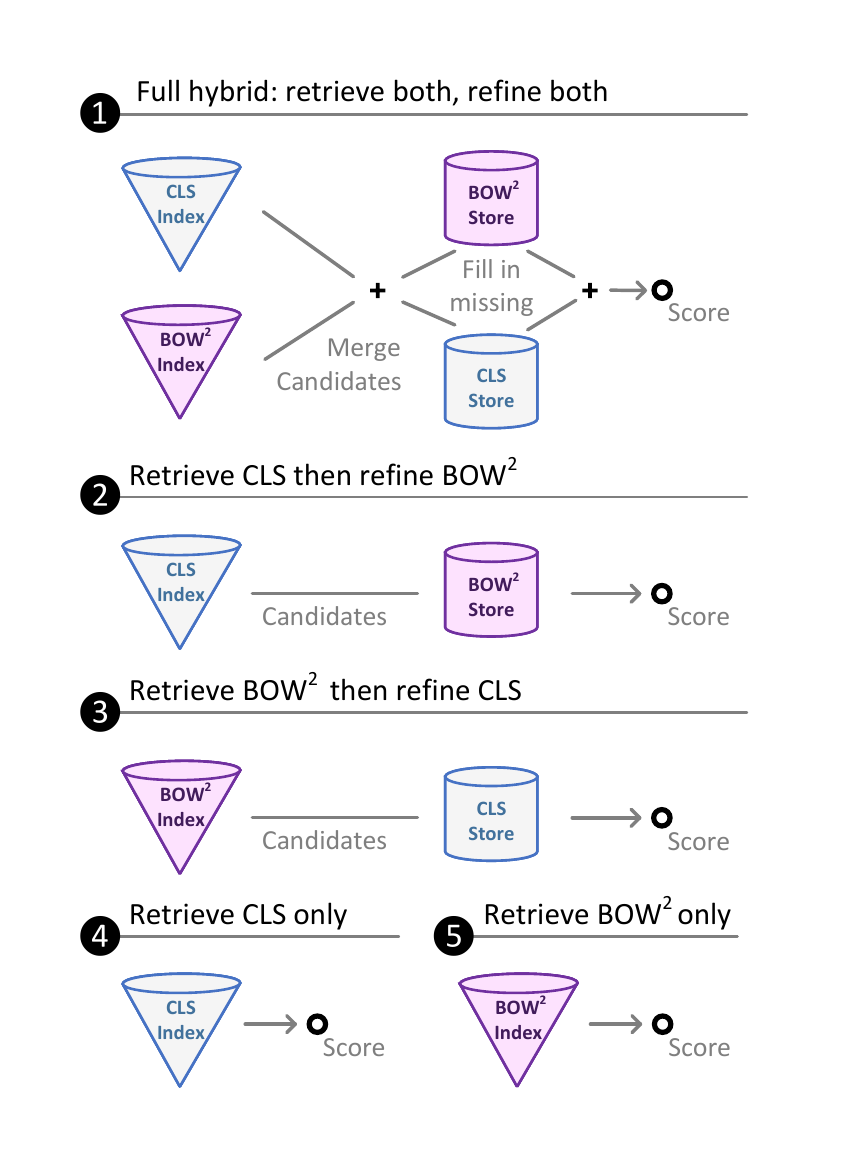}
    \centering
    \vspace{-0.3cm}
    \caption{The potential retrieval and refine workflows of ColBERTer at query time. Broadly categorized by: full hybrid (\ding{202}), single index, then refine with the other (\ding{203} + \ding{204}), or only one index for ablation purposes (\ding{205} + \ding{206}).}
    \label{fig:colberter_workflow}
    \vspace{-0.1cm}
\end{figure}

\subsection{Indexing and Query Workflow}
\label{sec:retrieval_workflows}

Once we have trained our ColBERTer model we need to decide how to deploy it into a wider retrieval workflow. ColBERTer's passage encoding can be fully pre-computed in an offline setting, which allows for low latency query-time retrieval. 

Previous works, such as COIL \cite{gao2021coil} or ColBERT \cite{khattab2020colbert} have already independently established many of the potential workflows. In addition to related approaches, we aim to give a holistic overview of all possible usage scenarios, including ablation studies to select the best method with the lowest possible complexity. We give a schematic overview over ColBERTer's retrieval workflows in Figure \ref{fig:colberter_workflow}. We assume that all passages have been encoded and stored accessibly by their id, as shown in Figure \ref{fig:colberter_architecture}. Each of the two storage categories can be transformed into an index structure for fast retrieval: the CLS index uses an (approximate) nearest neighbor index, while the BOW$^2$ index could use either a dense nearest neighbor index, or a classic inverted index (with activated exact matching component). 

Figure \ref{fig:colberter_workflow} \ding{202} shows how we can index both scoring components of ColBERTer and then consequently use the id-based storages to fill in missing scores for passages retrieved only by one of the indices. A similar workflow has already been explored by \citet{lin2021densifying} and \citet{gao2021coil}. Figure \ref{fig:colberter_workflow} \ding{203} \& \ding{204} utilize only one retrieval index, and fill up the missing scores from the complementary id-based storage. This approach works vice-versa for dense or sparse indices, and represents a clear complexity and additional index storage reduction, at the potential of lower recall. This approach is very much akin to a two stage retrieve and re-rank pipeline that has been studied extensively \cite{hofstaetter2020_crossarchitecture_kd,Hofstaetter2021_tasb_dense_retrieval,lin2020distilling}, but mostly with separate models for separate stages (which requires more indexing and encoding resources than our single ColBERTer model). Figure \ref{fig:colberter_workflow} \ding{205} \& \ding{206} represent ablation studies that only rely on one or the other index while disregarding the other scoring part.

Using different workflows may have large implications in terms of complexity, storage requirements, and effectiveness; therefore we always indicate the type of query workflow used (with the numbers given in Figure \ref{fig:colberter_workflow}) in our results section and provide a detailed ablation study in \sref{sec:source_of_effect}.

%% file: sections/5.experiment.design.tex
\vspace{-0.2cm}
\section{Experiment Design}

Our main training and inference dependencies are PyTorch~\cite{pytorch2017}, HuggingFace Transformers \cite{wolf2019huggingface}, and the nearest neighbor search library Faiss \cite{faiss2017}. For training we utilize TAS-Balanced \cite{Hofstaetter2021_tasb_dense_retrieval} retrieved negatives with BERT-based teacher ensemble scores \cite{hofstaetter2020_crossarchitecture_kd}. 

\vspace{-0.2cm}
\subsection{Passage Collection \& Query Sets}

For training and in-domain evaluation we use the MSMARCO-Passage (V1) collection \cite{msmarco16} with the sparsely-judged MSMARCO-DEV query set of 6,980 queries (used in the leaderboard) as well as the densely-judged 97 query set of combined TREC-DL '19 \cite{trec2019overview} and '20 \cite{trec2020overview}. For TREC graded relevance (0 = non relevant to 3 = perfect), we use the recommended binarization point of 2 for the recall metric. For out of domain experiments we refer to the ir\_datasets catalogue \cite{macavaney:sigir2021-irds} for collection specific information, as we utilized the standardized test sets for the collections. 

\vspace{-0.2cm}
\subsection{Parameter Settings}

As a basis for our model instances we use a 6-layer DistilBERT \cite{sanh2019distilbert} encoder as their initialization starting point. For our CLS vector we followed guidance by \citet{ma2021simple} to utilize 128 dimensions, as it provides enough capacity for retrieval. For token vectors, we study and present multiple parameter configurations between 32 and 1 dimension. We initialize models with final token output smaller than 32 with the checkpoint of the 32 dimensional model. The BOW$^2$ and CS components do not need any parameterization, other than using a Porter stemmer to aggregate unique words. These components only need to be parameterized in terms of the training loss influence $\alpha$'s. We thoroughly studied the robustness of the model to various configurations, as presented in \sref{sec:source_of_effect}.

%% file: sections/6.results.tex
\section{Results}
\label{sec:results}

We now address the individual research questions we introduced earlier. First, we study the source of ColBERTer's effectiveness, and under which conditions its components work; then we compare our results to related approaches; and finally we investigate the robustness of ColBERTer to out of domain collections.  

\subsection{Source of Effectiveness}
\label{sec:source_of_effect}

\begin{table}[t]
    \centering
    \caption{Analysis of different score aggregation and training methods for ColBERTer (2-way dim reduction only; CLS dim: 128, token dim: 32; Workflow \ding{203}) in terms of retrieval effectiveness. We compare refining full-retrieval results from ColBERTer's CLS vector (Own) and a TAS-Balanced retriever (TAS) with different multi-task loss weights $\alpha_b$ and $\alpha_{CLS}$. \textit{Highest Own in bold, lowest underlined.} 
    }
    \label{tab:agg_loss_ablation}
    \setlength\tabcolsep{3.2pt}
    \vspace{-0.3cm}
    \begin{tabular}{ccc!{\color{gray}\vrule}ll!{\color{lightgray}\vrule}ll!{\color{lightgray}\vrule}ll!{\color{lightgray}\vrule}ll}
       
       \toprule

       &\multicolumn{2}{c!{\color{lightgray}\vrule}}{\multirow{2}{*}{\textbf{Train Loss}}} &
       \multicolumn{4}{c!{\color{lightgray}\vrule}}{\textbf{TREC-DL'19+20}}&
       \multicolumn{4}{c}{\textbf{MSMARCO DEV}}\\

       &&&
       \multicolumn{2}{c!{\color{lightgray}\vrule}}{nDCG@10} & \multicolumn{2}{c!{\color{lightgray}\vrule}}{R@1K} & 
       \multicolumn{2}{c!{\color{lightgray}\vrule}}{MRR@10} & \multicolumn{2}{c}{R@1K} \\
       &$\alpha_b$&$\alpha_{CLS}$&  \textit{Own} & \textit{TAS} & \textit{Own} & \textit{TAS} & \textit{Own} & \textit{TAS} & \textit{Own} & \textit{TAS} \\
       \midrule
       \arrayrulecolor{lightgray}
        \multicolumn{6}{l}{\textbf{Fixed Score Aggregation}} \\
        \textcolor{gray}{1} & 1 & 0 & \underline{.684} & .740 & \underline{.565} & .861 & \underline{.336} & .386 & \underline{.773} & .978  \\
        \midrule
        \multicolumn{6}{l}{\textbf{Learned Score Aggregation}} \\
        \textcolor{gray}{2} & 1 & 0.1  & .726 & .728 & .783 & .861 & .384 & .386 & .952 & .978 \\
        \textcolor{gray}{3} & 1 & 0.2  & .728 & .731 & .794 & .861 & .384 & .385 & .957 & .978 \\
        \textcolor{gray}{4} & 1 & 0.5  & \textbf{.734} & .734 & \textbf{.807} & .861 & \textbf{.386} & .386 & .961 & .978 \\
        \textcolor{gray}{5} & 1 & 1.0  & .730 & .730 & .806 & .861 & .381 & .381 & \textbf{.962} & .978 \\

        \arrayrulecolor{black}
       \bottomrule
    \end{tabular}
    \vspace{-.5cm}
    
\end{table}

The ColBERTer model is essentially a complex multi-task architecture, even though these learning tasks work together to form an eventual end-to-end retrieval model. As the complexity of the architecture and training procedure grows, so must our understanding of the conditions in which the model works and where it fails. 

For our first investigation we aim to understand the interdependence between the CLS retrieval and token refinement capabilities. The related COIL architecture \cite{gao2021coil} aggregates their two-way dimension reduction in a sum without explicit weighting and feeds the sum through a single loss function. COIL uses both representation types (namely, CLS and token representations) as index, therefore it is not necessary for any of the components to work standalone. In the ColBERTer architecture, we want to support full retrieval capabilities of the CLS vector as candidate generator. If it fails, the quality of the refinement process does not matter anymore. Therefore, we study:

\RQone

\begin{table}[t]
    \centering
    \caption{Analysis of the bag of whole-words (BOW$^2$) and contextualized stopword training of ColBERTer (CLS dim: 128, token dim: 32; Workflow \ding{203}) using different multi-task loss parameters. 
    }
    \label{tab:stopword_ablation}
    \setlength\tabcolsep{2.0pt}
    \vspace{-0.3cm}
    \begin{tabular}{cccc!{\color{gray}\vrule}rr!{\color{lightgray}\vrule}rr!{\color{lightgray}\vrule}rr}
       
       \toprule

       &\multicolumn{3}{c!{\color{lightgray}\vrule}}{\textbf{Train Loss}} &
       \multicolumn{2}{c!{\color{lightgray}\vrule}}{\textbf{BOW$^2$ Vectors}} & 
       \multicolumn{2}{c!{\color{lightgray}\vrule}}{\textbf{DL'19+20}}&
       \multicolumn{2}{c}{\textbf{DEV}}\\

       &$\alpha_b$&$\alpha_{CLS}$&$\alpha_{CS}$&\# Saved & \% Stop. & \footnotesize{nDCG@10} & \footnotesize{R@1K} & \footnotesize{MRR@10} & \footnotesize{R@1K} \\
       \midrule
       \arrayrulecolor{lightgray}
        \multicolumn{6}{l}{\textbf{BOW$^2$ only}} \\
        \textcolor{gray}{1} & 1 & 0.5 & 0  & 43.2 & 0 \% & .731 & .815 & .387 & .963   \\
        \textcolor{gray}{2} & 1 & 0.1 & 0  & 43.2 & 0 \% & .736 & .806 & .387 & .960   \\
        \midrule
        \multicolumn{8}{l}{\textbf{BOW$^2$ + Contextualized Stopwords}} \\
        \textcolor{gray}{3} & 1 & 0.5 & 1    & 29.1 & 33 \% &  .731 & .811 & .382 & .965 \\
        \textcolor{gray}{4} & 1 & 0.1 & 1    & 27.8 & 36 \% &  .729 & .802 & .385 & .960  \\
        \textcolor{gray}{5} & 1 & 0.1 & 0.75 & 30.9 & 29 \% &  .730 & .805 & .387 & .961    \\
        \textcolor{gray}{6} & 1 & 0.1 & 0.5  & 36.7 & 15 \%  &  .725 & .806 & .387 & .962  \\

        \arrayrulecolor{black}
       \bottomrule
    \end{tabular}
    \vspace{-.5cm}
    
\end{table}

To isolate the CLS retrieval performance for workflow \ding{203} (dense CLS retrieval, followed by BOW$^2$ storage refinement) we compare different training and aggregation strategies with ColBERTer's CLS retrieval vs. re-ranking the candidate set retrieved by a standalone TAS-Balanced retriever in Table \ref{tab:agg_loss_ablation}. Using COIL's aggregation and training approach (by fixing $\sigma(\gamma) = 0.5 $ in Eq. \ref{eq:learned_score_agg} and setting $\alpha_{CLS} = 0$) we observe in line 1 that the CLS retrieval component fails substantially, compared to utilizing TAS-B. We postulate that this happens, as the token refinement component is more capable in determining relevance and therefore it dominates the changes in gradients, which minimizes the standalone capabilities of CLS retrieval. Now, with our proposed multi-task and learned score aggregation (lines 2-5) we observe much better CLS retrieval performance. While it still lacks a bit behind TAS-B in recall, these deficiencies do not manifest itself after refining the token scores for top-10 results in both TREC-DL and MSMARCO DEV. We selected the best performing setting in line 4 for our future experiments.

The next addition in our multi-task framework is the learned removal of stopwords. For this we add a third loss function $\mathcal{L}_{CS}$ that directly contradicts the objective of the main $\mathcal{L}_b$ loss. In Table \ref{tab:stopword_ablation} we show the tradeoff between retained BOW$^2$ vectors and effectiveness. In lines 1 \& 2 we see ColBERTer without the stopword components, here 43 vectors are saved with unique BOW$^2$ for MSMARCO (compared to 77 for all subword tokens). In lines 3 to 6 we study different loss weighting combinations with CS. While the ratio of removed stopwords is rather sensitive to the selected parameters, the effectivness values largely remain constant for lines 4 to 6. Based on the MRR value of the DEV set (with the smallest effectiveness change, but still 29 \% removed vectors) we select configuration 5 going forward, although we stress that our approach would also work well with the other settings, and cherry picking parameters is not needed. This setting reduces the number of vectors and therefore the storage requirement by a factor of $2.5$ compared to ColBERT, while keeping the same top-10 effectiveness (comparing Table \ref{tab:stopword_ablation} line 5 vs. Table \ref{tab:agg_loss_ablation} line 1 (TAS-B re-ranked).

A curious path for future work based on the results of Table \ref{tab:stopword_ablation} would be to use a conservative loss setting (such as line 6) that does not force a lot of the word removal gates to become zero (so as to not take away capacity from the loss surface for the ranking tasks), followed by the removal words with a non-zero (but still small) threshold during inference.

Following the ablation of training possibilities, we now turn towards the possible usage scenarios, as laid out in \sref{sec:retrieval_workflows}, and answer:

\RQthree

For this study we use ColBERTer with exact matching with 8 and 1 dimensions (Uni-ColBERTer) for BOW$^2$ vectors, as they are more likely to be used in an inverted index. The inverted index lookup is performed by our hashed id, with potential, but highly unlikely conflicts. Then we followed the approach proposed by COIL \cite{gao2021coil} and UniCOIL \cite{lin2021few} to compute dot products for all entries of a posting list for all exact matches between the query and the inverted index, followed by a summation per document, and subsequent sorting to receive a ranked list. 

In Table \ref{tab:query_modes} we show the results of our study grouped by the type of indexing and retrieval. For all indexing schemes, we use the same trained models. We start with an ablation of only one of the two scoring parts in line 1-4. Unsurprisingly, using only one of the scoring parts of ColBERTer results in an effectiveness reduction. What is surprising though, is the magnitude of the effectiveness drop of the inverted index only workflow \ding{206} compared both using only CLS retrieval (workflow \ding{205}) or refining the results with CLS scores (workflow \ding{202}). Continuing the results, in the single retrieval then refinement section in line 5-8, we see that once we combine both scoring parts, the underlying indexing approach matters very little at the top-10 effectiveness (comparing lines 5 \& 7, as well as lines 6 \& 8), only the reduced recall of the BOW$^2$ indexing is carried over. This a great result for the robustness of our system, it shows that it can be deployed in a variety of approaches, and practitioners are not locked into a specific retrieval approach. For example if one has made large investments in an inverted index system, they could build on these investments with Uni-ColBERTer. 

\begin{table}[t]
    \centering
    \caption{Analysis of the retrieval quality for different query-time retrieval and refinement workflows of ColBERTer with vector dimension of 8 or 1 (Uni-ColBERTer). \textit{nDCG and MRR at cutoff 10.}
    }
    \label{tab:query_modes}
    \setlength\tabcolsep{2.5pt}
    \vspace{-0.3cm}
    \begin{tabular}{ccl!{\color{lightgray}\vrule}l!{\color{lightgray}\vrule}rr!{\color{lightgray}\vrule}rr}
       
       \toprule

       &\multicolumn{2}{l!{\color{lightgray}\vrule}}{\multirow{2}{*}{\textbf{Workflow}}} &
       \multirow{2}{*}{\textbf{Model}} & 
       \multicolumn{2}{c!{\color{lightgray}\vrule}}{\textbf{DL'19+20}}&
       \multicolumn{2}{c}{\textbf{DEV}}\\

       && & &\footnotesize{nDCG} & \footnotesize{R@1K} & \footnotesize{MRR} & \footnotesize{R@1K} \\
       \midrule
       \arrayrulecolor{lightgray}
        \multicolumn{6}{l}{\textbf{Retrieval Only Ablation}} \\
        \textcolor{gray}{1} & \multirow{2}{*}{\ding{206}} & \multirow{2}{*}{BOW$^2$ only}  & ColBERTer (Dim8)  & .323 & .780 & .131 & .895   \\
        \textcolor{gray}{2} &  &                                                           & Uni-ColBERTer   & .280 & .758 & .122 & .880  \\
        \midrule
        \textcolor{gray}{3} & \multirow{2}{*}{\ding{205}} & \multirow{2}{*}{CLS only} & ColBERTer (Dim8)  & .669 & .795 & .326 & .958   \\
        \textcolor{gray}{4} & &                                                       & Uni-ColBERTer  & .674 & .789 & .328 & .958  \\

        \midrule
        \multicolumn{8}{l}{\textbf{Single Retrieval > Refinement}} \\
        \textcolor{gray}{5} & \multirow{2}{*}{\ding{203}} & \multirow{2}{*}{BOW$^2$ > CLS} & ColBERTer (Dim8)  & .730 & .780 & .373 & .895   \\
        \textcolor{gray}{6} & &                                                       & Uni-ColBERTer  & .724 & .673 & .369 & .880  \\
        \midrule
        \textcolor{gray}{7} & \multirow{2}{*}{\ding{204}} & \multirow{2}{*}{CLS > BOW$^2$} & ColBERTer (Dim8)  & .733 & .795 & .375 & .958   \\
        \textcolor{gray}{8} & &                                                       & Uni-ColBERTer  & .727 & .789 & .373 & .958  \\
        \midrule
        \multicolumn{8}{l}{\textbf{Hybrid Retrieval \& Refinement}} \\
        \textcolor{gray}{9} & \multirow{2}{*}{\ding{202}} & \multirow{2}{*}{Merge (\ding{203}+\ding{204})} & ColBERTer (Dim8)  & .734 & .873 & .376 & .981   \\
        \textcolor{gray}{10} & &                                                       & Uni-ColBERTer  & .728 & .865 & .374 & .979  \\

        \arrayrulecolor{black}
       \bottomrule
    \end{tabular}
    
\end{table}

\begin{figure}[t]
    \centering
    \includegraphics[width=0.49\textwidth,clip, trim=0.0cm 0.0cm 0.0cm 0.0cm]{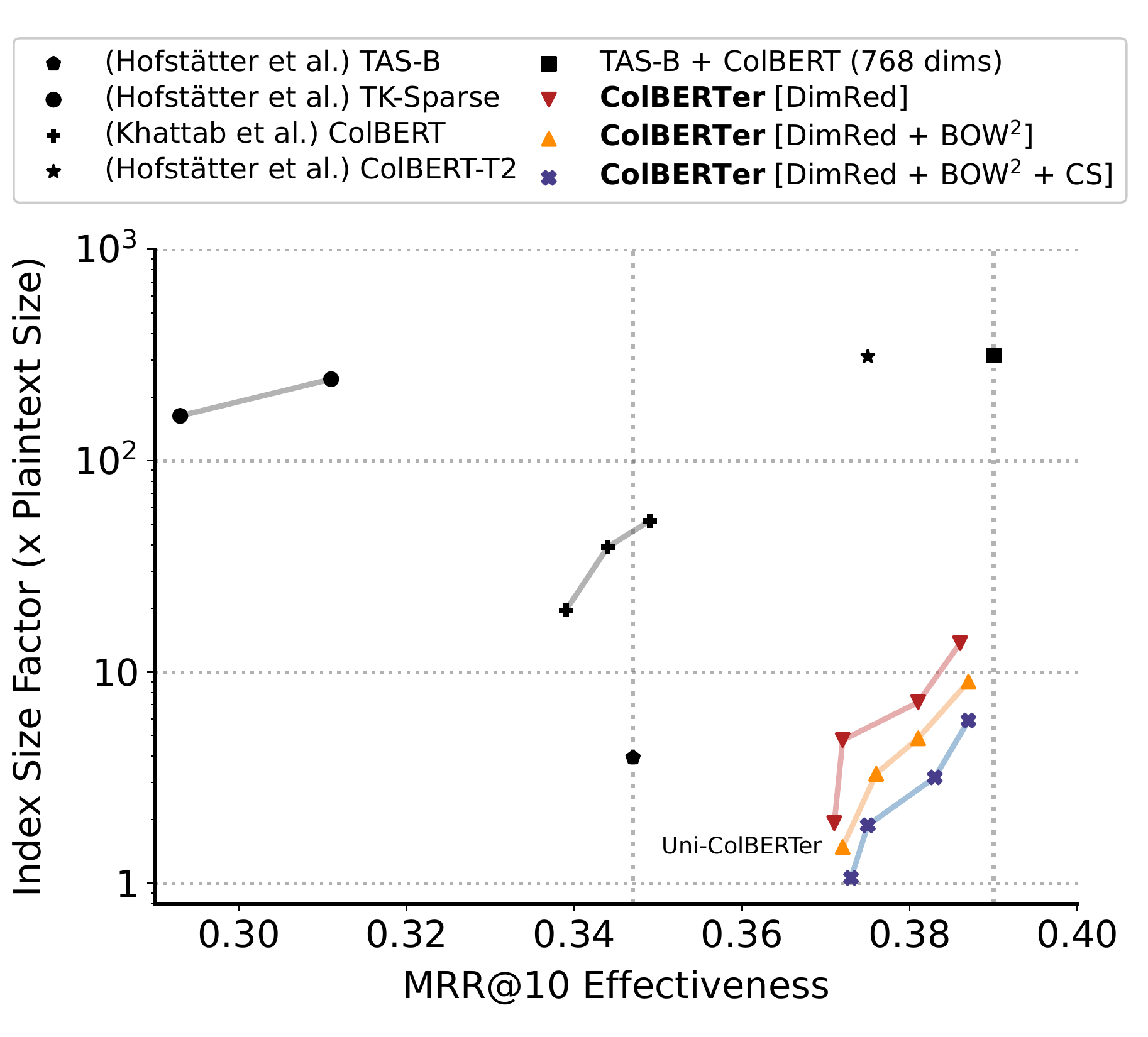}
    \vspace{-0.7cm}
  \caption{Tradeoff between storage requirements and effectiveness on MSMARCO Dev. \textit{Note the log scale of the y-axis.}}
  \label{fig:pareto-figure}
  \vspace{-0.5cm}
\end{figure}

Finally, we investigate a hybrid indexing workflow \ding{202}, where both index types generate candidates and all candidates are refined with the complimentary scoring part. We observe that the recall does increase compared to only one index, however, these improvements do not manifest themselves in the top-10 effectiveness. Here, the results are very close to the simpler workflows \ding{203} \& \ding{204}. Therefore, \textit{to keep it simple} we continue to use workflow \ding{203} and would suggest it as the primary way of using ColBERTer, if no previous investments make workflow \ding{204} more attractive. 

A general observation in the neural IR community is that more capacity in the number of vector dimensions usually leads to better results, albeit with diminishing returns. To see how our enhanced reduction fit into this assumption, we study:

\RQtwo

Our main concern of the enhanced reductions for the number of vectors introduced in ColBERTer is if they actually improve either or both effectiveness vs. costs, or if it would be better to keep the number of vectors the same and simply reduce the number of dimensions. In this case our work would be nullified. In Figure \ref{fig:pareto-figure} we show the tradeoff between storage requirements and effectiveness of our model configurations and closely related baselines.

First, we observe that the results of the single vector TAS-B \cite{Hofstaetter2021_tasb_dense_retrieval} and multi-vector staged pipeline of TAS-B + ColBERT (ours) form a corridor in which our ColBERTer results are expected to reside. Conforming with the expectations, all ColBERTer results are between the two in terms of effectiveness.

In Figure \ref{fig:pareto-figure} we display 3 ColBERTer reduction configurations for 32, 16, 8, and 1 (Uni-ColBERTer) token vector dimensions. Inside each configuration, we observe the usual pattern that more capacity improves effectiveness at the cost of increased storage requirements. Between configurations, we observe that removing half the vectors is more efficient and at the same time equal or even slightly improved effectiveness. Thus, using our enhanced reductions improves the Pareto frontier, compared to just reducing the dimensionality.

In the case of Uni-ColBERTer, there is no way of further reducing the dimensionality, so every removed vector enables previously unattainable efficiency gains. Our most efficient Uni-ColBERTer with all (BOW$^2$ and CS) reductions enabled reaches parity with the plaintext size it indexes. This includes the dense index which at 128 dimensions roughly takes up 2/3 of the total space.




\begin{table*}[t]
    \centering
    \caption{Comparing ColBERTers retrieval effectiveness to related approaches grouped by storage requirements. The storage factor refers to ratio of index to plaintext size of 3.05 GB. \textit{* indicates an estimation by us.}
    }
    \label{tab:related_work_comp}
    \setlength\tabcolsep{2.4pt}
    \vspace{-0.3cm}
    \begin{tabular}{cll!{\color{lightgray}\vrule}ll!{\color{lightgray}\vrule}r!{\color{lightgray}\vrule}c!{\color{lightgray}\vrule}ll!{\color{lightgray}\vrule}ll!{\color{lightgray}\vrule}ll}
      %
      %
       \toprule

       &\multicolumn{2}{c!{\color{lightgray}\vrule}}{\textbf{Model}} &
       \multicolumn{2}{c!{\color{lightgray}\vrule}}{\textbf{Storage}} &
       \textbf{Query} &
       \textbf{Interpret.} &
       \multicolumn{2}{c!{\color{lightgray}\vrule}}{\textbf{TREC-DL'19}}&
       \multicolumn{2}{c!{\color{lightgray}\vrule}}{\textbf{TREC-DL'20}}&
       \multicolumn{2}{c}{\textbf{DEV}}\\
       && & Total & Factor & \textbf{Latency} & \textbf{Ranking} & \footnotesize{nDCG@10}  & \footnotesize{R@1K} & \footnotesize{nDCG@10} & \footnotesize{R@1K} & \footnotesize{MRR@10} & \footnotesize{R@1K} \\

       \midrule
       \arrayrulecolor{lightgray}
       \multicolumn{12}{l}{\textbf{Low Storage Systems (max. 2x Factor)}} \\

       \textcolor{gray}{1} & \cite{mackenzie2021wacky} & BM25 (PISA) & 0.7 GB & $\times$ 0.2  & 8 ms & \cmark &  .501  & .739 & .475 & .806 &  .194 & .868 \\
       \midrule

       \textcolor{gray}{2} & \cite{zhan2021jointly} & JPQ           &  0.8 GB & $\times$ 0.3 & 90 ms &  \xmark & .677 & -- & -- & -- & .341 & --       \\
       \textcolor{gray}{3} & \cite{lin2021few} & UniCOIL-Tok           &    N/A  & N/A  & N/A & \cmark & -- & -- & -- & -- & .315 & -- \\
       \textcolor{gray}{4} & \cite{mackenzie2021wacky} & UniCOIL-Tok (+docT5query)   & 1.4 GB & $\times$ 0.5 & 37 ms & \cmark & -- & -- & -- & -- & .352 & -- \\
       \textcolor{gray}{5} & \cite{formal2021splade,mackenzie2021wacky} & SPLADEv2 (PISA)  & 4.3 GB & $\times$ 1.4 & 220 ms & \xmark & .729 & -- & -- & -- & .369 & \textbf{.979} \\
       \textcolor{gray}{6} & \cite{lin2021densifying} & DSR-SPLADE + Dense-CLS (Dim 128)  & 5 GB & $\times$ 1.6 & 32 ms & \xmark & .709 & -- & .673 & -- & .344 & --         \\
       
       \midrule
       \textcolor{gray}{7} &  & Uni-ColBERTer (Dim 1) & 3.3 GB & $\times$ 1.1 & 55 ms & \cmark & .727 & .761 & .726 & .812 & .373 & .958  \\
       \textcolor{gray}{8} &  & ColBERTer w. EM (Dim 8) & 5.8 GB & $\times$ 1.9 & 55 ms & \cmark & \textbf{.732} & \textbf{.764} & \textbf{.734} & \textbf{.819} & \textbf{.375} & .958  \\

       \arrayrulecolor{black}
       \midrule
       \arrayrulecolor{lightgray}
       \multicolumn{12}{l}{\textbf{Higher Storage Systems}} \\

       \textcolor{gray}{9} & \cite{gao2021coil} & COIL (Dim 128, 8)  & 12.5 GB* & $\times$ 4.1*  & 21 ms & \cmark & .694 & -- & -- & -- & .347 & .956  \\
       \textcolor{gray}{10} & \cite{gao2021coil} & COIL (Dim 768, 32)   & 54.7 GB* & $\times$ 17.9 & 41 ms & \cmark & .704 & -- & -- & -- & .355 & \textbf{.963}  \\
       \textcolor{gray}{11} & \cite{lin2021densifying} & DSR-SPLADE + Dense-CLS (Dim 256)  & 11 GB & $\times$ 3.6 & 34 ms & \xmark & .711 & -- & .678 & -- & .348 & --         \\
       \textcolor{gray}{12} & \cite{lin2021few,lin2020distilling} & TCT-ColBERTv2 + UniCOIL (+dT5q)   & 14.4 GB* & $\times$ 4.7* & 110 ms & \cmark & -- & -- & -- & -- & .378 & --  \\
       \midrule
       \textcolor{gray}{13} &  & ColBERTer (Dim 16) & 9.9 GB  & $\times$ 3.2 & 51 ms & \cmark & .726 & .782 & .719 & .829 & .383 & .961  \\
       

       \textcolor{gray}{14} &  & ColBERTer (Dim 32) & 18.8 GB & $\times$ 6.2 & 51 ms  & \cmark & \textbf{.727} & \textbf{.781} & \textbf{.733} & \textbf{.825} & \textbf{.387} & .961  \\

        \arrayrulecolor{black}
       \bottomrule
    \vspace{-.7cm}
    \end{tabular}
\end{table*}

\subsection{Comparing to Related Work}

We want to emphasize that it becomes increasingly hard to contrast neural retrieval models and make conclusive statements about "SOTA" (State-of-the-art). This is because there are numerous factors at play for the effectiveness including training data sampling, distillation, generational training, and even hardware setups. At the same time it is highly important to compare systems not only by their effectiveness, but factor in the efficiency as well, to avoid misleading claims. We believe it is important to show that we do not observe substantial differences in effectiveness compared to other systems of similar efficiency and that small deviations of effectiveness should not strongly impact our overall assessment -- even if those small differences come out in our favor. With that in mind, we study:

\RQfive

In Table \ref{tab:related_work_comp} we group models by our main efficiency focus: the storage requirements, measured as the factor of the indexed plaintext size. 

\paragraph{\textbf{Low Storage Systems}}
We find that ColBERTer does improve on the existing Pareto frontier compared to other approaches, especially for cases with low storage footprint. Our Uni-ColBERTer (line 7) especially outperforms previous single-dimension token encoding approaches, while at the same time offering improved transparency and making it easier to showcase model scores with mappings to whole words. However, we also observe that we could further improve the dense retrieval component with a technique similar to JPQ \cite{zhan2021jointly} (line 2) to further reduce our storage footprint.

\paragraph{\textbf{Higher Storage Systems}}
At first glance in this section, we see that while 32 dimensions per token does not sound much, the resulting increase of the total storage requirement is staggering. ColBERTer outperforms similarly sized architectures as well, however a fair comparison becomes more difficult than in the low storage systems, as the absolute size differences become much larger. Another curious observation is that larger ColBERTer models (lines 13 \& 14) seem to be slightly faster than our smaller instances (lines 7 \& 8). We believe this is due to the fact that we utilize non-optimized python code to lookup the top-1000 token storage memory locations per query, which takes 10ms for ColBERTer without exact matching and 15 ms for ColBERTer with exact matching as there we need to access 2 locations per passage (one for the values and one for the ids). There is definitely potential for strong optimizations in the future or production implementation work for this aspect.

\subsection{Out-of-Domain Robustness}

In this section we evaluate the zero-shot performance of our ColBERTer architecture, when it is applied on retrieval collections from domains outside the training data to answer:

\RQsix

\begin{figure*}[t]
  \centering
  \begin{subfigure}[t]{0.4\textwidth}
    \centering
    \includegraphics[height=5.3cm,clip, trim=0.2cm 0.2cm 0.2cm 0.2cm]{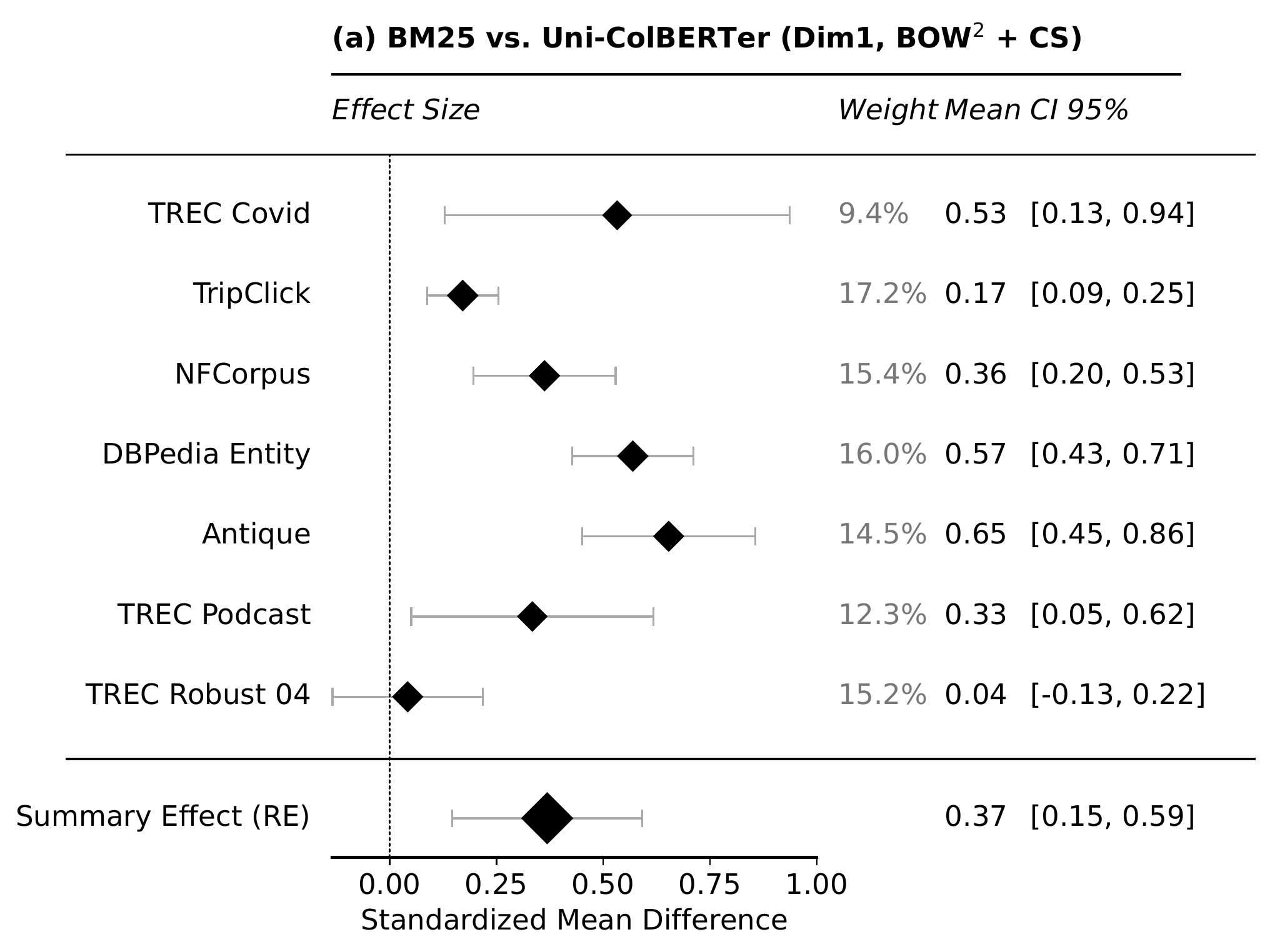}
  \end{subfigure}%
  \begin{subfigure}[t]{0.3\textwidth}
    \centering
    \includegraphics[height=5.3cm,clip, trim=5.1cm 0.2cm 0.2cm 0.2cm]{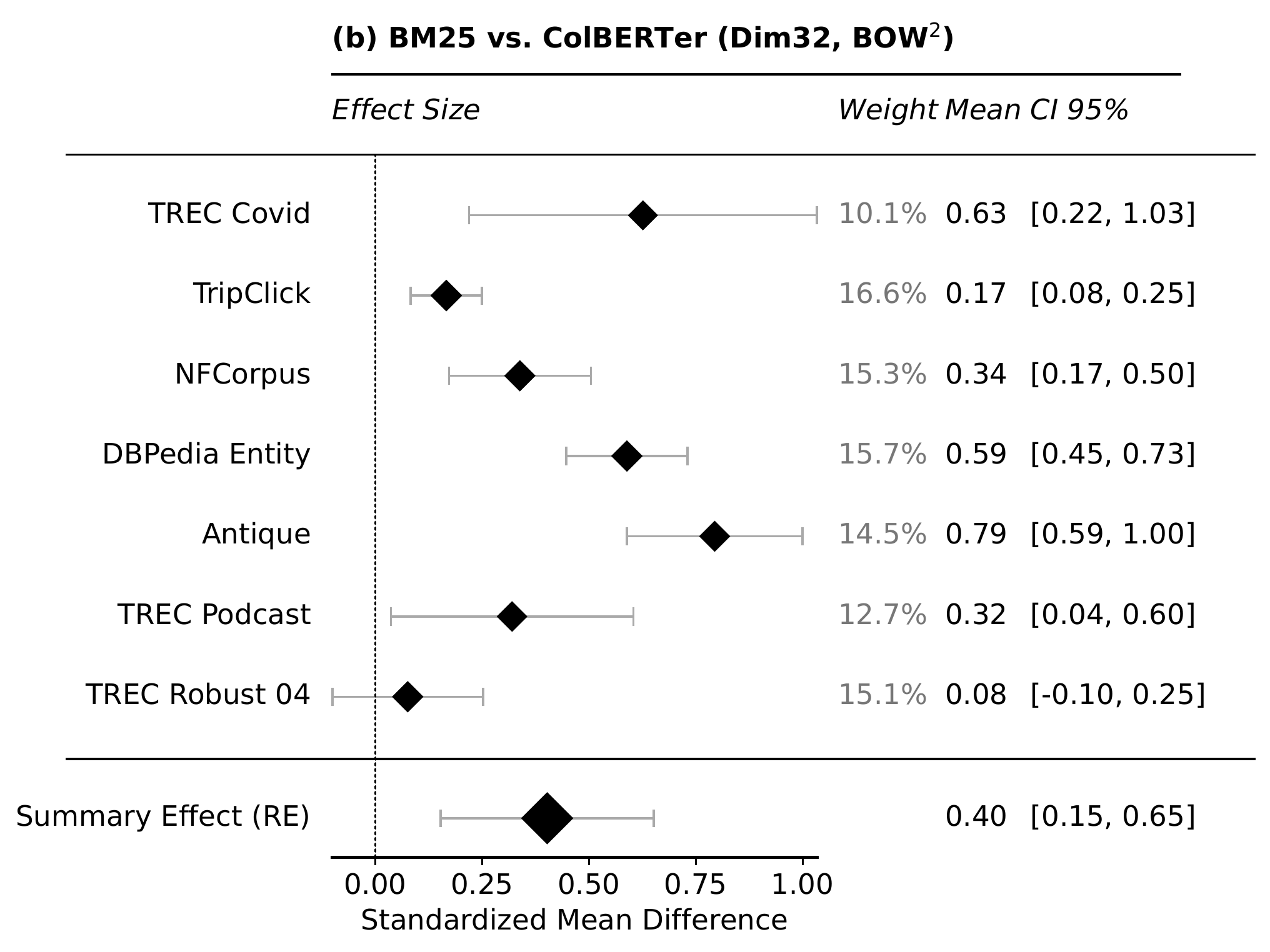}
  \end{subfigure}%
    \begin{subfigure}[t]{0.3\textwidth}
    \centering
    \includegraphics[height=5.3cm,clip, trim=5.1cm 0.2cm 0.2cm 0.2cm]{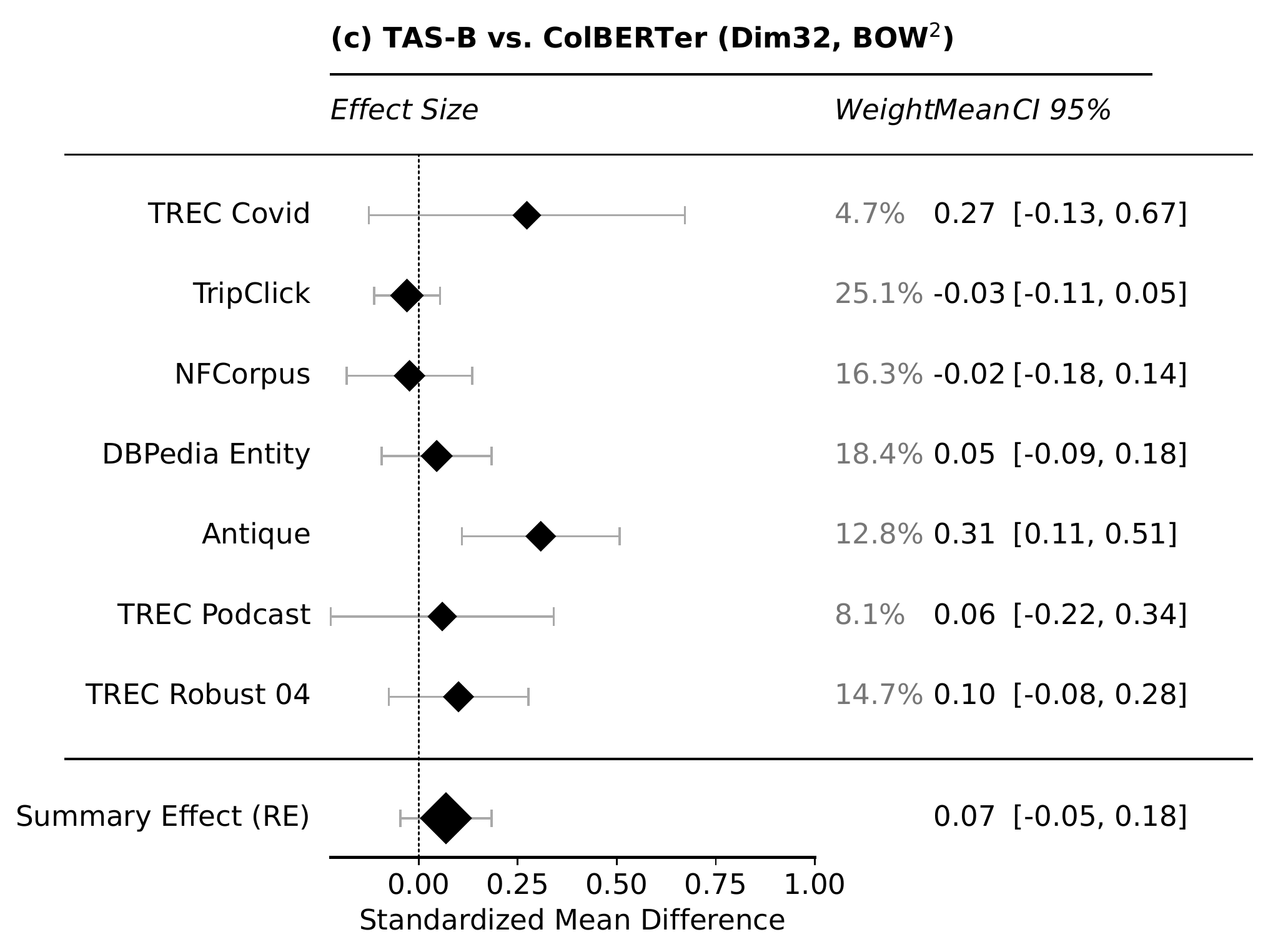}
  \end{subfigure}%
  \caption{Effect size based evaluation of ColBERTer's zero-shot out of domain robustness. We compare three pairings between the control vs. the treatment retrieval method. The comparison is dependent on the effect size of each collection and the mean NDCG@10 differences are standardized with the effect size. The confidence intervals are plotted as interval around the standardized mean difference \ding{117}. The Summary Effect of the treatment is computed with the Random-Effect (RE) model, here we see an overall significant improvement for ColBERTer (Dim1 and Dim32) to BM25.}
  \label{fig:ood-robustness-effect-size}
\end{figure*}

Our main aim is to present an analysis grounded in robust evaluation \cite{voorhees2001philosophy,zovel1998reliability} that does not fall for common problematic shortcuts in IR evaluation like influence of effect sizes \cite{fuhr2018commonmistakes, webber2008stat_power}, relying on too shallow pooled collections \cite{arabzadeh2021shallow,webber2009poolbias,lu2016effect}, not accounting for pool bias in old collections \cite{buckley2004retrieval, Tetsuya2007dashJ, sakai2008poolbias}, and aggregating metrics over different collections which are not comparable \cite{soboroff2018meta}. We first describe our evaluation methodology and then discuss our results presented in Figure \ref{fig:ood-robustness-effect-size}.

\paragraph{\textbf{Methodology}}

We selected seven datasets from the \texttt{ir\_datasets} catalogue \cite{macavaney:sigir2021-irds}: Bio medical (TREC Covid \cite{Wang2020Cord19,Voorhees2020TrecCovid}, TripClick \cite{Rekabsaz2021TripClick}, NFCorpus \cite{Boteva2016Nfcorpus}), Entity centric (DBPedia Entity \cite{Hasibi2017DBpediaEntityVA}), informal language (Antique \cite{hashemi2020antique}, TREC Podcast \cite{jones2021trec}), news cables (TREC Robust 04 \cite{Voorhees2004Robust}). The datasets are not based on web collections, have at least $50$ queries, and importantly contain judgements from both relevant and non-relevant categories. Three datasets are also part of the BEIR \cite{thakur2021beir} catalogue. We choose not to use other datasets from BEIR, as they do not contain non-relevant judgements, which makes it impossible to conduct pooling bias corrections. 

We follow \citet{Tetsuya2007dashJ} to correct our metric measurements for pool bias by observing only measuring effectiveness on judged passages (which means removing all retrieved passages that are not judged and then re-assign the ranks of the remaining ones). This is in contrast with the default assumption that non-judged passages are not relevant (which of course favors methods that have been part of the pooling process). Additionally, we follow \citet{soboroff2018meta} to utilize an effect size analysis that is popular in medicine and social sciences. \citet{soboroff2018meta} proposed to use this effect size as meta analysis tool to be able to compare statistical significance across different retrieval collections. In this work we combine the evaluation approaches of \citet{Tetsuya2007dashJ} and \citet{soboroff2018meta} for the first time to receive strong confidence in our results and analysis.

We take the standardized mean difference (SMD) in nDCG@10 score between a baseline model and our model as the effect. 
Besides the variability within different collections, we assume, as \citet{soboroff2018meta}, a between collection heterogeneity. 
Following \citet{soboroff2018meta}, we use a random-effect model to estimate the summary effect of our model and each individual effect's contribution, i.e., weight. We use the DerSimonian and Laird estimate \cite{dersimonian2015meta} to obtain the between collection variance.
We illustrate the outcome of our meta-analysis as forest plots.
Diamonds \ding{117} show the effect in each collection and, in turn, in summary.
Each effect is accompanied by its $95\%$ confidence interval -- the grey line.
The dotted vertical line marks \textit{null effect}, i.e., zero SMD in nDCG@10 score between our model and the compared baseline.
A confidence interval crossing the null effect line indicates that the corresponding effect is statistically not significant; in all other cases, it contains the actual effect of our model $95\%$ of the time. 

As baseline, we utilize BM25 as implemented by Pyserini \cite{lin2021pyserini}. We apply our models, trained on MSMARCO, end-to-end in a zero-shot fashion with our default settings for retrieval. We compare a ColBERTer version with $32$ token dimensions, as well as Uni-ColBERTer with a single token dimension and exact matching prior. 

\paragraph{\textbf{Discussion}}

We illustrate the effect of using Uni-ColBERTer instead of BM25 within the different collections and the corresponding summary effect in Figure \ref{fig:ood-robustness-effect-size}a.
Compared to the retrospective approach of hypothesis testing with p-values, confidence intervals are predictive \cite{soboroff2018meta}.
Considering the TripClick collection, for example, we expect the effect to be between .09 and .25 95\% of the time, indicating that we can detect the effect size of .17 SMD at the given confidence level and underlining the significant effectiveness gains using Uni-ColBERTer over BM25. Only on TREC Robust 04 is the small improved difference inside a 95\% confidence interval.
Overall, by judging the summary effect in Figure \ref{fig:ood-robustness-effect-size}a, we expect that choosing Uni-ColBERTer over BM25 consistently and significantly improves effectiveness.
Similarly, considering Figure \ref{fig:ood-robustness-effect-size}b, we expect ColBERTer (Dim32) to consistently and significantly outperform BM25.
However, comparing the summary effects in Figure \ref{fig:ood-robustness-effect-size}a and Figure \ref{fig:ood-robustness-effect-size}b, we expect Uni-ColBERTer and ColBERTer (Dim32) to behave similarly if run against BM25, suggesting to use the more efficient model.
We also compare our model to a more effective neural dense retriever TAS-B \cite{Hofstaetter2021_tasb_dense_retrieval}, which has been shown to work effectively in an out of domain setting \cite{thakur2021beir}.
We report the effect of using ColBERTer (Dim32) vs. TAS-B in Figure \ref{fig:ood-robustness-effect-size}c. Here, we see a much less clear image than in the other two cases. Most collections overlap inside the 95\% CI, including the summary effect model, which must lead us to the conclusion that these models are equally effective. Only the Antique collection is significantly improved by ColBERTer. The TREC Covid collection is a curious case -- looking at absolute numbers, one would easily assume a substantial improvement, however, because it only evaluates 50 queries the confidence interval is very wide. Finally, what does this mean for a deployment decision of ColBERTer vs. TAS-B? We need to consider other aspects, such as transparency. We would argue ColBERTer increases transparency over TAS-B as laid out in this paper, and at the same time it does not show a single collection with significantly worse results, therefore we would select it.

%% file: sections/7.conclusion.tex
\section{Conclusion}

In this paper, we proposed ColBERTer, an efficient and effective retrieval model that improves the storage efficiency, the retrieval complexity, and the interpretability of the ColBERT architecture along the effectiveness Pareto frontier. To this end, ColBERTer learns whole-word representations that exclude contextualized stopwords, yielding 2.5$\times$ fewer vectors than ColBERT while supporting user-friendly query--document scoring patterns at the level of whole words. ColBERTer also uses a multi-task, multi-stage training objective---as well as an optional lexical matching component---that together enable it to aggressively reduce the vector dimension to 1. Extensive empirical evaluation shows that ColBERTer is highly effective on MS MARCO and TREC-DL and highly robust out of domain, while demonstrating highly-competitive storage efficiency with prior dense and sparse models.

%% file: main.bbl

\begin{thebibliography}{60}


\ifx \showCODEN    \undefined \def \showCODEN     #1{\unskip}     \fi
\ifx \showDOI      \undefined \def \showDOI       #1{#1}\fi
\ifx \showISBNx    \undefined \def \showISBNx     #1{\unskip}     \fi
\ifx \showISBNxiii \undefined \def \showISBNxiii  #1{\unskip}     \fi
\ifx \showISSN     \undefined \def \showISSN      #1{\unskip}     \fi
\ifx \showLCCN     \undefined \def \showLCCN      #1{\unskip}     \fi
\ifx \shownote     \undefined \def \shownote      #1{#1}          \fi
\ifx \showarticletitle \undefined \def \showarticletitle #1{#1}   \fi
\ifx \showURL      \undefined \def \showURL       {\relax}        \fi
\providecommand\bibfield[2]{#2}
\providecommand\bibinfo[2]{#2}
\providecommand\natexlab[1]{#1}
\providecommand\showeprint[2][]{arXiv:#2}

\bibitem[\protect\citeauthoryear{??}{der}{2015}]%
        {dersimonian2015meta}
 \bibinfo{year}{2015}\natexlab{}.
\newblock \showarticletitle{Meta-analysis in clinical trials revisited}.
\newblock \bibinfo{journal}{\emph{Contemporary Clinical Trials}}
  \bibinfo{volume}{45} (\bibinfo{year}{2015}), \bibinfo{pages}{139--145}.
\newblock
\showISSN{1551-7144}
\urldef\tempurl%
\url{https://doi.org/10.1016/j.cct.2015.09.002}
\showDOI{\tempurl}
\newblock
\shownote{10th Anniversary Special Issue.}


\bibitem[\protect\citeauthoryear{Arabzadeh, Vtyurina, Yan, and
  Clarke}{Arabzadeh et~al\mbox{.}}{2021}]%
        {arabzadeh2021shallow}
\bibfield{author}{\bibinfo{person}{Negar Arabzadeh}, \bibinfo{person}{Alexandra
  Vtyurina}, \bibinfo{person}{Xinyi Yan}, {and} \bibinfo{person}{Charles~LA
  Clarke}.} \bibinfo{year}{2021}\natexlab{}.
\newblock \showarticletitle{Shallow pooling for sparse labels}.
\newblock \bibinfo{journal}{\emph{arXiv preprint arXiv:2109.00062}}
  (\bibinfo{year}{2021}).
\newblock


\bibitem[\protect\citeauthoryear{Bajaj, Campos, Craswell, Deng, Gao, Liu,
  Majumder, Mcnamara, Mitra, and Nguyen}{Bajaj et~al\mbox{.}}{2016}]%
        {msmarco16}
\bibfield{author}{\bibinfo{person}{Payal Bajaj}, \bibinfo{person}{Daniel
  Campos}, \bibinfo{person}{Nick Craswell}, \bibinfo{person}{Li Deng},
  \bibinfo{person}{Jianfeng Gao}, \bibinfo{person}{Xiaodong Liu},
  \bibinfo{person}{Rangan Majumder}, \bibinfo{person}{Andrew Mcnamara},
  \bibinfo{person}{Bhaskar Mitra}, {and} \bibinfo{person}{Tri Nguyen}.}
  \bibinfo{year}{2016}\natexlab{}.
\newblock \showarticletitle{{MS MARCO : A Human Generated MAchine Reading
  COmprehension Dataset}}. In \bibinfo{booktitle}{\emph{Proc. of NIPS}}.
\newblock


\bibitem[\protect\citeauthoryear{Boteva, Gholipour, Sokolov, and
  Riezler}{Boteva et~al\mbox{.}}{2016}]%
        {Boteva2016Nfcorpus}
\bibfield{author}{\bibinfo{person}{Vera Boteva}, \bibinfo{person}{Demian
  Gholipour}, \bibinfo{person}{Artem Sokolov}, {and} \bibinfo{person}{Stefan
  Riezler}.} \bibinfo{year}{2016}\natexlab{}.
\newblock \showarticletitle{A Full-Text Learning to Rank Dataset for Medical
  Information Retrieval}. In \bibinfo{booktitle}{\emph{Proceedings of the
  European Conference on Information Retrieval ({ECIR})}} (Padova, Italy).
  \bibinfo{publisher}{Springer}.
\newblock


\bibitem[\protect\citeauthoryear{Buckley and Voorhees}{Buckley and
  Voorhees}{2004}]%
        {buckley2004retrieval}
\bibfield{author}{\bibinfo{person}{Chris Buckley} {and}
  \bibinfo{person}{Ellen~M Voorhees}.} \bibinfo{year}{2004}\natexlab{}.
\newblock \showarticletitle{Retrieval evaluation with incomplete information}.
  In \bibinfo{booktitle}{\emph{Proceedings of the 27th annual international ACM
  SIGIR conference on Research and development in information retrieval}}.
  \bibinfo{pages}{25--32}.
\newblock


\bibitem[\protect\citeauthoryear{Castillo}{Castillo}{2019}]%
        {castello2019fairranking}
\bibfield{author}{\bibinfo{person}{Carlos Castillo}.}
  \bibinfo{year}{2019}\natexlab{}.
\newblock \showarticletitle{Fairness and Transparency in Ranking}.
\newblock \bibinfo{journal}{\emph{SIGIR Forum}} \bibinfo{volume}{52},
  \bibinfo{number}{2} (\bibinfo{date}{jan} \bibinfo{year}{2019}),
  \bibinfo{pages}{64–71}.
\newblock
\showISSN{0163-5840}
\urldef\tempurl%
\url{https://doi.org/10.1145/3308774.3308783}
\showDOI{\tempurl}


\bibitem[\protect\citeauthoryear{Craswell, Mitra, Yilmaz, and Campos}{Craswell
  et~al\mbox{.}}{2019}]%
        {trec2019overview}
\bibfield{author}{\bibinfo{person}{Nick Craswell}, \bibinfo{person}{Bhaskar
  Mitra}, \bibinfo{person}{Emine Yilmaz}, {and} \bibinfo{person}{Daniel
  Campos}.} \bibinfo{year}{2019}\natexlab{}.
\newblock \showarticletitle{Overview of the TREC 2019 deep learning track}. In
  \bibinfo{booktitle}{\emph{TREC}}.
\newblock


\bibitem[\protect\citeauthoryear{Craswell, Mitra, Yilmaz, and Campos}{Craswell
  et~al\mbox{.}}{2020}]%
        {trec2020overview}
\bibfield{author}{\bibinfo{person}{Nick Craswell}, \bibinfo{person}{Bhaskar
  Mitra}, \bibinfo{person}{Emine Yilmaz}, {and} \bibinfo{person}{Daniel
  Campos}.} \bibinfo{year}{2020}\natexlab{}.
\newblock \showarticletitle{Overview of the TREC 2020 Deep Learning Track}. In
  \bibinfo{booktitle}{\emph{TREC}}.
\newblock


\bibitem[\protect\citeauthoryear{Devlin, Chang, Lee, and Toutanova}{Devlin
  et~al\mbox{.}}{2018}]%
        {devlin2018bert}
\bibfield{author}{\bibinfo{person}{Jacob Devlin}, \bibinfo{person}{Ming-Wei
  Chang}, \bibinfo{person}{Kenton Lee}, {and} \bibinfo{person}{Kristina
  Toutanova}.} \bibinfo{year}{2018}\natexlab{}.
\newblock \showarticletitle{BERT: Pre-training of Deep Bidirectional
  Transformers for Language Understanding}.
\newblock \bibinfo{journal}{\emph{arXiv preprint arXiv:1810.04805}}
  (\bibinfo{year}{2018}).
\newblock


\bibitem[\protect\citeauthoryear{Formal, Lassance, Piwowarski, and
  Clinchant}{Formal et~al\mbox{.}}{2021}]%
        {formal2021splade}
\bibfield{author}{\bibinfo{person}{Thibault Formal}, \bibinfo{person}{Carlos
  Lassance}, \bibinfo{person}{Benjamin Piwowarski}, {and}
  \bibinfo{person}{St{\'e}phane Clinchant}.} \bibinfo{year}{2021}\natexlab{}.
\newblock \showarticletitle{SPLADE v2: Sparse lexical and expansion model for
  information retrieval}.
\newblock \bibinfo{journal}{\emph{arXiv preprint arXiv:2109.10086}}
  (\bibinfo{year}{2021}).
\newblock


\bibitem[\protect\citeauthoryear{Fuhr}{Fuhr}{2018}]%
        {fuhr2018commonmistakes}
\bibfield{author}{\bibinfo{person}{Norbert Fuhr}.}
  \bibinfo{year}{2018}\natexlab{}.
\newblock \showarticletitle{Some Common Mistakes In IR Evaluation, And How They
  Can Be Avoided}.
\newblock \bibinfo{journal}{\emph{SIGIR Forum}} \bibinfo{volume}{51},
  \bibinfo{number}{3} (\bibinfo{date}{feb} \bibinfo{year}{2018}),
  \bibinfo{pages}{32–41}.
\newblock
\showISSN{0163-5840}
\urldef\tempurl%
\url{https://doi.org/10.1145/3190580.3190586}
\showDOI{\tempurl}


\bibitem[\protect\citeauthoryear{Gao, Dai, and Callan}{Gao
  et~al\mbox{.}}{2021}]%
        {gao2021coil}
\bibfield{author}{\bibinfo{person}{Luyu Gao}, \bibinfo{person}{Zhuyun Dai},
  {and} \bibinfo{person}{Jamie Callan}.} \bibinfo{year}{2021}\natexlab{}.
\newblock \showarticletitle{COIL: Revisit Exact Lexical Match in Information
  Retrieval with Contextualized Inverted List}.
\newblock \bibinfo{journal}{\emph{arXiv preprint arXiv:2104.07186}}
  (\bibinfo{year}{2021}).
\newblock


\bibitem[\protect\citeauthoryear{Hashemi, Aliannejadi, Zamani, and
  Croft}{Hashemi et~al\mbox{.}}{2020}]%
        {hashemi2020antique}
\bibfield{author}{\bibinfo{person}{Helia Hashemi}, \bibinfo{person}{Mohammad
  Aliannejadi}, \bibinfo{person}{Hamed Zamani}, {and} \bibinfo{person}{W~Bruce
  Croft}.} \bibinfo{year}{2020}\natexlab{}.
\newblock \showarticletitle{ANTIQUE: A non-factoid question answering
  benchmark}. In \bibinfo{booktitle}{\emph{Proc. of ECIR}}.
\newblock


\bibitem[\protect\citeauthoryear{Hasibi, Nikolaev, Xiong, Balog, Bratsberg,
  Kotov, and Callan}{Hasibi et~al\mbox{.}}{2017}]%
        {Hasibi2017DBpediaEntityVA}
\bibfield{author}{\bibinfo{person}{Faegheh Hasibi}, \bibinfo{person}{Fedor
  Nikolaev}, \bibinfo{person}{Chenyan Xiong}, \bibinfo{person}{K. Balog},
  \bibinfo{person}{S.~E. Bratsberg}, \bibinfo{person}{Alexander Kotov}, {and}
  \bibinfo{person}{J. Callan}.} \bibinfo{year}{2017}\natexlab{}.
\newblock \showarticletitle{DBpedia-Entity v2: A Test Collection for Entity
  Search}.
\newblock \bibinfo{journal}{\emph{Proceedings of the 40th International ACM
  SIGIR Conference on Research and Development in Information Retrieval}}
  (\bibinfo{year}{2017}).
\newblock


\bibitem[\protect\citeauthoryear{Hofst{\"a}tter, Althammer, Schr{\"o}der,
  Sertkan, and Hanbury}{Hofst{\"a}tter et~al\mbox{.}}{2020a}]%
        {hofstaetter2020_crossarchitecture_kd}
\bibfield{author}{\bibinfo{person}{Sebastian Hofst{\"a}tter},
  \bibinfo{person}{Sophia Althammer}, \bibinfo{person}{Michael Schr{\"o}der},
  \bibinfo{person}{Mete Sertkan}, {and} \bibinfo{person}{Allan Hanbury}.}
  \bibinfo{year}{2020}\natexlab{a}.
\newblock \showarticletitle{Improving Efficient Neural Ranking Models with
  Cross-Architecture Knowledge Distillation}.
\newblock \bibinfo{journal}{\emph{arXiv:2010.02666}} (\bibinfo{year}{2020}).
\newblock


\bibitem[\protect\citeauthoryear{Hofst{\"a}tter, Lin, Yang, Lin, and
  Hanbury}{Hofst{\"a}tter et~al\mbox{.}}{2021}]%
        {Hofstaetter2021_tasb_dense_retrieval}
\bibfield{author}{\bibinfo{person}{Sebastian Hofst{\"a}tter},
  \bibinfo{person}{Sheng-Chieh Lin}, \bibinfo{person}{Jheng-Hong Yang},
  \bibinfo{person}{Jimmy Lin}, {and} \bibinfo{person}{Allan Hanbury}.}
  \bibinfo{year}{2021}\natexlab{}.
\newblock \showarticletitle{{Efficiently Teaching an Effective Dense Retriever
  with Balanced Topic Aware Sampling}}. In
  \bibinfo{booktitle}{\emph{Proceedings of the 44rd International ACM SIGIR
  Conference on Research and Development in Information Retrieval (SIGIR
  '21)}}.
\newblock


\bibitem[\protect\citeauthoryear{Hofst{\"a}tter, Lipani, Zlabinger, and
  Hanbury}{Hofst{\"a}tter et~al\mbox{.}}{2020b}]%
        {Hofstaetter2020_cikm}
\bibfield{author}{\bibinfo{person}{Sebastian Hofst{\"a}tter},
  \bibinfo{person}{Aldo Lipani}, \bibinfo{person}{Markus Zlabinger}, {and}
  \bibinfo{person}{Allan Hanbury}.} \bibinfo{year}{2020}\natexlab{b}.
\newblock \showarticletitle{{Learning to Re-Rank with Contextualized
  Stopwords}}. In \bibinfo{booktitle}{\emph{Proc. of CIKM}}.
\newblock


\bibitem[\protect\citeauthoryear{Hofst{\"a}tter, Zlabinger, and
  Hanbury}{Hofst{\"a}tter et~al\mbox{.}}{2020c}]%
        {Hofstaetter2020_ecai}
\bibfield{author}{\bibinfo{person}{Sebastian Hofst{\"a}tter},
  \bibinfo{person}{Markus Zlabinger}, {and} \bibinfo{person}{Allan Hanbury}.}
  \bibinfo{year}{2020}\natexlab{c}.
\newblock \showarticletitle{{Interpretable \& Time-Budget-Constrained
  Contextualization for Re-Ranking}}. In \bibinfo{booktitle}{\emph{Proc. of
  ECAI}}.
\newblock


\bibitem[\protect\citeauthoryear{Hofst{\"a}tter, Zlabinger, and
  Hanbury}{Hofst{\"a}tter et~al\mbox{.}}{2020d}]%
        {Hofstaetter2020_ecir}
\bibfield{author}{\bibinfo{person}{Sebastian Hofst{\"a}tter},
  \bibinfo{person}{Markus Zlabinger}, {and} \bibinfo{person}{Allan Hanbury}.}
  \bibinfo{year}{2020}\natexlab{d}.
\newblock \showarticletitle{{Neural-IR-Explorer: A Content-Focused Tool to
  Explore Neural Re-Ranking Results}}. In \bibinfo{booktitle}{\emph{Proc. of
  ECIR}}.
\newblock


\bibitem[\protect\citeauthoryear{Humeau, Shuster, Lachaux, and Weston}{Humeau
  et~al\mbox{.}}{2020}]%
        {humeau2020polyencoders}
\bibfield{author}{\bibinfo{person}{Samuel Humeau}, \bibinfo{person}{Kurt
  Shuster}, \bibinfo{person}{Marie-Anne Lachaux}, {and} \bibinfo{person}{Jason
  Weston}.} \bibinfo{year}{2020}\natexlab{}.
\newblock \showarticletitle{Poly-encoders: Transformer Architectures and
  Pre-training Strategies for Fast and Accurate Multi-sentence Scoring}.
\newblock \bibinfo{journal}{\emph{1Proceedings of the International Conference
  on Learning Representations (ICLR) 2020}}.
\newblock


\bibitem[\protect\citeauthoryear{Ji, Shao, and Yang}{Ji et~al\mbox{.}}{2019}]%
        {ji2019efficient}
\bibfield{author}{\bibinfo{person}{Shiyu Ji}, \bibinfo{person}{Jinjin Shao},
  {and} \bibinfo{person}{Tao Yang}.} \bibinfo{year}{2019}\natexlab{}.
\newblock \showarticletitle{Efficient Interaction-based Neural Ranking with
  Locality Sensitive Hashing}. In \bibinfo{booktitle}{\emph{Proc of. WWW}}.
\newblock


\bibitem[\protect\citeauthoryear{Johnson, Douze, and J{\'e}gou}{Johnson
  et~al\mbox{.}}{2017}]%
        {faiss2017}
\bibfield{author}{\bibinfo{person}{Jeff Johnson}, \bibinfo{person}{Matthijs
  Douze}, {and} \bibinfo{person}{Herv{\'e} J{\'e}gou}.}
  \bibinfo{year}{2017}\natexlab{}.
\newblock \showarticletitle{Billion-Scale Similarity Search with {GPUs}}.
\newblock \bibinfo{journal}{\emph{arXiv:1702.08734}} (\bibinfo{year}{2017}).
\newblock


\bibitem[\protect\citeauthoryear{Jones, Carterette, Clifton, Eskevich, Jones,
  Karlgren, Pappu, Reddy, and Yu}{Jones et~al\mbox{.}}{2021}]%
        {jones2021trec}
\bibfield{author}{\bibinfo{person}{Rosie Jones}, \bibinfo{person}{Ben
  Carterette}, \bibinfo{person}{Ann Clifton}, \bibinfo{person}{Maria Eskevich},
  \bibinfo{person}{Gareth~JF Jones}, \bibinfo{person}{Jussi Karlgren},
  \bibinfo{person}{Aasish Pappu}, \bibinfo{person}{Sravana Reddy}, {and}
  \bibinfo{person}{Yongze Yu}.} \bibinfo{year}{2021}\natexlab{}.
\newblock \showarticletitle{Trec 2020 podcasts track overview}.
\newblock \bibinfo{journal}{\emph{arXiv preprint arXiv:2103.15953}}
  (\bibinfo{year}{2021}).
\newblock


\bibitem[\protect\citeauthoryear{Khattab and Zaharia}{Khattab and
  Zaharia}{2020}]%
        {khattab2020colbert}
\bibfield{author}{\bibinfo{person}{Omar Khattab} {and} \bibinfo{person}{Matei
  Zaharia}.} \bibinfo{year}{2020}\natexlab{}.
\newblock \showarticletitle{ColBERT: Efficient and Effective Passage Search via
  Contextualized Late Interaction over BERT}. In
  \bibinfo{booktitle}{\emph{Proc. of SIGIR}}.
\newblock


\bibitem[\protect\citeauthoryear{Lassance, Maachou, Park, and
  Clinchant}{Lassance et~al\mbox{.}}{2021}]%
        {lassance2021study}
\bibfield{author}{\bibinfo{person}{Carlos Lassance}, \bibinfo{person}{Maroua
  Maachou}, \bibinfo{person}{Joohee Park}, {and} \bibinfo{person}{Stéphane
  Clinchant}.} \bibinfo{year}{2021}\natexlab{}.
\newblock \bibinfo{title}{A Study on Token Pruning for ColBERT}.
\newblock
\newblock
\showeprint[arxiv]{2112.06540}~[cs.IR]


\bibitem[\protect\citeauthoryear{Lee, Sung, Kang, and Chen}{Lee
  et~al\mbox{.}}{2020}]%
        {lee2020learning}
\bibfield{author}{\bibinfo{person}{Jinhyuk Lee}, \bibinfo{person}{Mujeen Sung},
  \bibinfo{person}{Jaewoo Kang}, {and} \bibinfo{person}{Danqi Chen}.}
  \bibinfo{year}{2020}\natexlab{}.
\newblock \showarticletitle{Learning dense representations of phrases at
  scale}.
\newblock \bibinfo{journal}{\emph{arXiv preprint arXiv:2012.12624}}
  (\bibinfo{year}{2020}).
\newblock


\bibitem[\protect\citeauthoryear{Lewis, Oğuz, Xiong, Petroni, tau Yih, and
  Riedel}{Lewis et~al\mbox{.}}{2021}]%
        {lewis2021boosted}
\bibfield{author}{\bibinfo{person}{Patrick Lewis}, \bibinfo{person}{Barlas
  Oğuz}, \bibinfo{person}{Wenhan Xiong}, \bibinfo{person}{Fabio Petroni},
  \bibinfo{person}{Wen tau Yih}, {and} \bibinfo{person}{Sebastian Riedel}.}
  \bibinfo{year}{2021}\natexlab{}.
\newblock \showarticletitle{Boosted Dense Retriever}.
\newblock \bibinfo{journal}{\emph{arXiv preprint arXiv:2112.07771}}
  (\bibinfo{year}{2021}).
\newblock


\bibitem[\protect\citeauthoryear{Lin and Ma}{Lin and Ma}{2021}]%
        {lin2021few}
\bibfield{author}{\bibinfo{person}{Jimmy Lin} {and} \bibinfo{person}{Xueguang
  Ma}.} \bibinfo{year}{2021}\natexlab{}.
\newblock \showarticletitle{A few brief notes on deepimpact, coil, and a
  conceptual framework for information retrieval techniques}.
\newblock \bibinfo{journal}{\emph{arXiv preprint arXiv:2106.14807}}
  (\bibinfo{year}{2021}).
\newblock


\bibitem[\protect\citeauthoryear{Lin, Ma, Lin, Yang, Pradeep, and Nogueira}{Lin
  et~al\mbox{.}}{2021}]%
        {lin2021pyserini}
\bibfield{author}{\bibinfo{person}{Jimmy Lin}, \bibinfo{person}{Xueguang Ma},
  \bibinfo{person}{Sheng-Chieh Lin}, \bibinfo{person}{Jheng-Hong Yang},
  \bibinfo{person}{Ronak Pradeep}, {and} \bibinfo{person}{Rodrigo Nogueira}.}
  \bibinfo{year}{2021}\natexlab{}.
\newblock \showarticletitle{Pyserini: A Python Toolkit for Reproducible
  Information Retrieval Research with Sparse and Dense Representations}. In
  \bibinfo{booktitle}{\emph{Proc. of SIGIR}}.
\newblock


\bibitem[\protect\citeauthoryear{Lin and Lin}{Lin and Lin}{2021}]%
        {lin2021densifying}
\bibfield{author}{\bibinfo{person}{Sheng-Chieh Lin} {and}
  \bibinfo{person}{Jimmy Lin}.} \bibinfo{year}{2021}\natexlab{}.
\newblock \showarticletitle{Densifying Sparse Representations for Passage
  Retrieval by Representational Slicing}.
\newblock \bibinfo{journal}{\emph{arXiv preprint arXiv:2112.04666}}
  (\bibinfo{year}{2021}).
\newblock


\bibitem[\protect\citeauthoryear{Lin, Yang, and Lin}{Lin et~al\mbox{.}}{2020}]%
        {lin2020distilling}
\bibfield{author}{\bibinfo{person}{Sheng-Chieh Lin},
  \bibinfo{person}{Jheng-Hong Yang}, {and} \bibinfo{person}{Jimmy Lin}.}
  \bibinfo{year}{2020}\natexlab{}.
\newblock \showarticletitle{Distilling Dense Representations for Ranking using
  Tightly-Coupled Teachers}.
\newblock \bibinfo{journal}{\emph{arXiv:2010.11386}} (\bibinfo{year}{2020}).
\newblock


\bibitem[\protect\citeauthoryear{Lu, Jiao, and Zhang}{Lu et~al\mbox{.}}{2020}]%
        {lu2020twinbert}
\bibfield{author}{\bibinfo{person}{Wenhao Lu}, \bibinfo{person}{Jian Jiao},
  {and} \bibinfo{person}{Ruofei Zhang}.} \bibinfo{year}{2020}\natexlab{}.
\newblock \showarticletitle{TwinBERT: Distilling Knowledge to Twin-Structured
  BERT Models for Efficient Retrieval}.
\newblock \bibinfo{journal}{\emph{arXiv:2002.06275}} (\bibinfo{year}{2020}).
\newblock


\bibitem[\protect\citeauthoryear{Lu, Moffat, and Culpepper}{Lu
  et~al\mbox{.}}{2016}]%
        {lu2016effect}
\bibfield{author}{\bibinfo{person}{Xiaolu Lu}, \bibinfo{person}{Alistair
  Moffat}, {and} \bibinfo{person}{J~Shane Culpepper}.}
  \bibinfo{year}{2016}\natexlab{}.
\newblock \showarticletitle{The effect of pooling and evaluation depth on IR
  metrics}.
\newblock \bibinfo{journal}{\emph{Information Retrieval Journal}}
  \bibinfo{volume}{19}, \bibinfo{number}{4} (\bibinfo{year}{2016}),
  \bibinfo{pages}{416--445}.
\newblock


\bibitem[\protect\citeauthoryear{Luan, Eisenstein, Toutanova, and Collins}{Luan
  et~al\mbox{.}}{2020}]%
        {luan2020sparse}
\bibfield{author}{\bibinfo{person}{Yi Luan}, \bibinfo{person}{Jacob
  Eisenstein}, \bibinfo{person}{Kristina Toutanova}, {and}
  \bibinfo{person}{Michael Collins}.} \bibinfo{year}{2020}\natexlab{}.
\newblock \showarticletitle{Sparse, Dense, and Attentional Representations for
  Text Retrieval}.
\newblock \bibinfo{journal}{\emph{arXiv preprint arXiv:2005.00181}}
  (\bibinfo{year}{2020}).
\newblock


\bibitem[\protect\citeauthoryear{Luan, Eisenstein, Toutanova, and Collins}{Luan
  et~al\mbox{.}}{2021}]%
        {luan-etal-2021-sparsemebert}
\bibfield{author}{\bibinfo{person}{Yi Luan}, \bibinfo{person}{Jacob
  Eisenstein}, \bibinfo{person}{Kristina Toutanova}, {and}
  \bibinfo{person}{Michael Collins}.} \bibinfo{year}{2021}\natexlab{}.
\newblock \showarticletitle{Sparse, Dense, and Attentional Representations for
  Text Retrieval}.
\newblock \bibinfo{journal}{\emph{Transactions of the Association for
  Computational Linguistics}}  \bibinfo{volume}{9} (\bibinfo{year}{2021}),
  \bibinfo{pages}{329--345}.
\newblock
\urldef\tempurl%
\url{https://doi.org/10.1162/tacl_a_00369}
\showDOI{\tempurl}


\bibitem[\protect\citeauthoryear{Ma, Li, Sun, Xin, and Lin}{Ma
  et~al\mbox{.}}{2021}]%
        {ma2021simple}
\bibfield{author}{\bibinfo{person}{Xueguang Ma}, \bibinfo{person}{Minghan Li},
  \bibinfo{person}{Kai Sun}, \bibinfo{person}{Ji Xin}, {and}
  \bibinfo{person}{Jimmy Lin}.} \bibinfo{year}{2021}\natexlab{}.
\newblock \showarticletitle{Simple and Effective Unsupervised Redundancy
  Elimination to Compress Dense Vectors for Passage Retrieval}. In
  \bibinfo{booktitle}{\emph{Proceedings of the 2021 Conference on Empirical
  Methods in Natural Language Processing}}.
\newblock


\bibitem[\protect\citeauthoryear{MacAvaney, Yates, Feldman, Downey, Cohan, and
  Goharian}{MacAvaney et~al\mbox{.}}{2021}]%
        {macavaney:sigir2021-irds}
\bibfield{author}{\bibinfo{person}{Sean MacAvaney}, \bibinfo{person}{Andrew
  Yates}, \bibinfo{person}{Sergey Feldman}, \bibinfo{person}{Doug Downey},
  \bibinfo{person}{Arman Cohan}, {and} \bibinfo{person}{Nazli Goharian}.}
  \bibinfo{year}{2021}\natexlab{}.
\newblock \showarticletitle{Simplified Data Wrangling with ir\_datasets}. In
  \bibinfo{booktitle}{\emph{SIGIR}}.
\newblock


\bibitem[\protect\citeauthoryear{Mackenzie, Trotman, and Lin}{Mackenzie
  et~al\mbox{.}}{2021}]%
        {mackenzie2021wacky}
\bibfield{author}{\bibinfo{person}{Joel Mackenzie}, \bibinfo{person}{Andrew
  Trotman}, {and} \bibinfo{person}{Jimmy Lin}.}
  \bibinfo{year}{2021}\natexlab{}.
\newblock \showarticletitle{Wacky weights in learned sparse representations and
  the revenge of score-at-a-time query evaluation}.
\newblock \bibinfo{journal}{\emph{arXiv preprint arXiv:2110.11540}}
  (\bibinfo{year}{2021}).
\newblock


\bibitem[\protect\citeauthoryear{Paszke, Gross, Chintala, Chanan,
  et~al\mbox{.}}{Paszke et~al\mbox{.}}{2017}]%
        {pytorch2017}
\bibfield{author}{\bibinfo{person}{Adam Paszke}, \bibinfo{person}{Sam Gross},
  \bibinfo{person}{Soumith Chintala}, \bibinfo{person}{Gregory Chanan},
  {et~al\mbox{.}}} \bibinfo{year}{2017}\natexlab{}.
\newblock \showarticletitle{Automatic differentiation in PyTorch}. In
  \bibinfo{booktitle}{\emph{NIPS-W}}.
\newblock


\bibitem[\protect\citeauthoryear{Rekabsaz, Lesota, Schedl, Brassey, and
  Eickhoff}{Rekabsaz et~al\mbox{.}}{2021}]%
        {Rekabsaz2021TripClick}
\bibfield{author}{\bibinfo{person}{Navid Rekabsaz}, \bibinfo{person}{Oleg
  Lesota}, \bibinfo{person}{Markus Schedl}, \bibinfo{person}{Jon Brassey},
  {and} \bibinfo{person}{Carsten Eickhoff}.} \bibinfo{year}{2021}\natexlab{}.
\newblock \showarticletitle{TripClick: The Log Files of a Large Health Web
  Search Engine}. In \bibinfo{booktitle}{\emph{SIGIR}}.
\newblock


\bibitem[\protect\citeauthoryear{Sakai}{Sakai}{2007}]%
        {Tetsuya2007dashJ}
\bibfield{author}{\bibinfo{person}{Tetsuya Sakai}.}
  \bibinfo{year}{2007}\natexlab{}.
\newblock \showarticletitle{Alternatives to Bpref}. In
  \bibinfo{booktitle}{\emph{Proc. of SIGIR}}.
\newblock


\bibitem[\protect\citeauthoryear{Sakai}{Sakai}{2008}]%
        {sakai2008poolbias}
\bibfield{author}{\bibinfo{person}{Tetsuya Sakai}.}
  \bibinfo{year}{2008}\natexlab{}.
\newblock \showarticletitle{Comparing Metrics across TREC and NTCIR: The
  Robustness to System Bias}. In \bibinfo{booktitle}{\emph{Proc. of CIKM}}.
\newblock


\bibitem[\protect\citeauthoryear{Sanh, Debut, Chaumond, and Wolf}{Sanh
  et~al\mbox{.}}{2019}]%
        {sanh2019distilbert}
\bibfield{author}{\bibinfo{person}{Victor Sanh}, \bibinfo{person}{Lysandre
  Debut}, \bibinfo{person}{Julien Chaumond}, {and} \bibinfo{person}{Thomas
  Wolf}.} \bibinfo{year}{2019}\natexlab{}.
\newblock \showarticletitle{DistilBERT, a distilled version of BERT: smaller,
  faster, cheaper and lighter}.
\newblock \bibinfo{journal}{\emph{arXiv preprint arXiv:1910.01108}}
  (\bibinfo{year}{2019}).
\newblock


\bibitem[\protect\citeauthoryear{Santhanam, Khattab, Saad-Falcon, Potts, and
  Zaharia}{Santhanam et~al\mbox{.}}{2021}]%
        {santhanam2021colbertv2}
\bibfield{author}{\bibinfo{person}{Keshav Santhanam}, \bibinfo{person}{Omar
  Khattab}, \bibinfo{person}{Jon Saad-Falcon}, \bibinfo{person}{Christopher
  Potts}, {and} \bibinfo{person}{Matei Zaharia}.}
  \bibinfo{year}{2021}\natexlab{}.
\newblock \showarticletitle{ColBERTv2: Effective and Efficient Retrieval via
  Lightweight Late Interaction}.
\newblock \bibinfo{journal}{\emph{arXiv preprint arXiv:2112.01488}}
  (\bibinfo{year}{2021}).
\newblock


\bibitem[\protect\citeauthoryear{Schuster and Nakajima}{Schuster and
  Nakajima}{2012}]%
        {schuster2012japanese}
\bibfield{author}{\bibinfo{person}{Mike Schuster} {and}
  \bibinfo{person}{Kaisuke Nakajima}.} \bibinfo{year}{2012}\natexlab{}.
\newblock \showarticletitle{Japanese and korean voice search}. In
  \bibinfo{booktitle}{\emph{2012 IEEE International Conference on Acoustics,
  Speech and Signal Processing (ICASSP)}}. IEEE, \bibinfo{pages}{5149--5152}.
\newblock


\bibitem[\protect\citeauthoryear{Soboroff}{Soboroff}{2018}]%
        {soboroff2018meta}
\bibfield{author}{\bibinfo{person}{Ian Soboroff}.}
  \bibinfo{year}{2018}\natexlab{}.
\newblock \showarticletitle{Meta-Analysis for Retrieval Experiments Involving
  Multiple Test Collections}. In \bibinfo{booktitle}{\emph{Proc. of CIKM}}.
\newblock


\bibitem[\protect\citeauthoryear{Thakur, Reimers, Rücklé, Srivastava, and
  Gurevych}{Thakur et~al\mbox{.}}{2021}]%
        {thakur2021beir}
\bibfield{author}{\bibinfo{person}{Nandan Thakur}, \bibinfo{person}{Nils
  Reimers}, \bibinfo{person}{Andreas Rücklé}, \bibinfo{person}{Abhishek
  Srivastava}, {and} \bibinfo{person}{Iryna Gurevych}.}
  \bibinfo{year}{2021}\natexlab{}.
\newblock \bibinfo{title}{BEIR: A Heterogenous Benchmark for Zero-shot
  Evaluation of Information Retrieval Models}.
\newblock
\newblock
\showeprint[arxiv]{2104.08663}~[cs.IR]


\bibitem[\protect\citeauthoryear{Tonellotto and Macdonald}{Tonellotto and
  Macdonald}{2021}]%
        {tonelloto2021embeddingpruning}
\bibfield{author}{\bibinfo{person}{Nicola Tonellotto} {and}
  \bibinfo{person}{Craig Macdonald}.} \bibinfo{year}{2021}\natexlab{}.
\newblock \bibinfo{booktitle}{\emph{Query Embedding Pruning for Dense
  Retrieval}}.
\newblock \bibinfo{publisher}{Association for Computing Machinery},
  \bibinfo{address}{New York, NY, USA}, \bibinfo{pages}{3453–3457}.
\newblock
\showISBNx{9781450384469}
\urldef\tempurl%
\url{https://doi.org/10.1145/3459637.3482162}
\showURL{%
\tempurl}


\bibitem[\protect\citeauthoryear{Voorhees}{Voorhees}{2004}]%
        {Voorhees2004Robust}
\bibfield{author}{\bibinfo{person}{Ellen Voorhees}.}
  \bibinfo{year}{2004}\natexlab{}.
\newblock \showarticletitle{Overview of the TREC 2004 Robust Retrieval Track}.
  In \bibinfo{booktitle}{\emph{TREC}}.
\newblock


\bibitem[\protect\citeauthoryear{Voorhees, Alam, Bedrick, Demner-Fushman,
  Hersh, Lo, Roberts, Soboroff, and Wang}{Voorhees et~al\mbox{.}}{2020}]%
        {Voorhees2020TrecCovid}
\bibfield{author}{\bibinfo{person}{E. Voorhees}, \bibinfo{person}{Tasmeer
  Alam}, \bibinfo{person}{Steven Bedrick}, \bibinfo{person}{Dina
  Demner-Fushman}, \bibinfo{person}{W. Hersh}, \bibinfo{person}{Kyle Lo},
  \bibinfo{person}{Kirk Roberts}, \bibinfo{person}{I. Soboroff}, {and}
  \bibinfo{person}{Lucy~Lu Wang}.} \bibinfo{year}{2020}\natexlab{}.
\newblock \showarticletitle{TREC-COVID: Constructing a Pandemic Information
  Retrieval Test Collection}.
\newblock \bibinfo{journal}{\emph{ArXiv}}  \bibinfo{volume}{abs/2005.04474}
  (\bibinfo{year}{2020}).
\newblock


\bibitem[\protect\citeauthoryear{Voorhees}{Voorhees}{2001}]%
        {voorhees2001philosophy}
\bibfield{author}{\bibinfo{person}{Ellen~M Voorhees}.}
  \bibinfo{year}{2001}\natexlab{}.
\newblock \showarticletitle{The philosophy of information retrieval
  evaluation}. In \bibinfo{booktitle}{\emph{Workshop of the cross-language
  evaluation forum for european languages}}. Springer,
  \bibinfo{pages}{355--370}.
\newblock


\bibitem[\protect\citeauthoryear{Wang, Lo, Chandrasekhar, Reas, Yang, Eide,
  Funk, Kinney, Liu, Merrill, Mooney, Murdick, Rishi, Sheehan, Shen, Stilson,
  Wade, Wang, Wilhelm, Xie, Raymond, Weld, Etzioni, and Kohlmeier}{Wang
  et~al\mbox{.}}{2020}]%
        {Wang2020Cord19}
\bibfield{author}{\bibinfo{person}{Lucy~Lu Wang}, \bibinfo{person}{Kyle Lo},
  \bibinfo{person}{Yoganand Chandrasekhar}, \bibinfo{person}{Russell Reas},
  \bibinfo{person}{Jiangjiang Yang}, \bibinfo{person}{Darrin Eide},
  \bibinfo{person}{K. Funk}, \bibinfo{person}{Rodney~Michael Kinney},
  \bibinfo{person}{Ziyang Liu}, \bibinfo{person}{W. Merrill},
  \bibinfo{person}{P. Mooney}, \bibinfo{person}{D. Murdick},
  \bibinfo{person}{Devvret Rishi}, \bibinfo{person}{Jerry Sheehan},
  \bibinfo{person}{Zhihong Shen}, \bibinfo{person}{B. Stilson},
  \bibinfo{person}{A. Wade}, \bibinfo{person}{K. Wang},
  \bibinfo{person}{Christopher Wilhelm}, \bibinfo{person}{Boya Xie},
  \bibinfo{person}{D. Raymond}, \bibinfo{person}{Daniel~S. Weld},
  \bibinfo{person}{Oren Etzioni}, {and} \bibinfo{person}{Sebastian Kohlmeier}.}
  \bibinfo{year}{2020}\natexlab{}.
\newblock \showarticletitle{CORD-19: The Covid-19 Open Research Dataset}.
\newblock \bibinfo{journal}{\emph{ArXiv}} (\bibinfo{year}{2020}).
\newblock


\bibitem[\protect\citeauthoryear{Webber, Moffat, and Zobel}{Webber
  et~al\mbox{.}}{2008}]%
        {webber2008stat_power}
\bibfield{author}{\bibinfo{person}{William Webber}, \bibinfo{person}{Alistair
  Moffat}, {and} \bibinfo{person}{Justin Zobel}.}
  \bibinfo{year}{2008}\natexlab{}.
\newblock \showarticletitle{Statistical Power in Retrieval Experimentation}. In
  \bibinfo{booktitle}{\emph{Proc. of CIKM}}.
\newblock


\bibitem[\protect\citeauthoryear{Webber and Park}{Webber and Park}{2009}]%
        {webber2009poolbias}
\bibfield{author}{\bibinfo{person}{William Webber} {and}
  \bibinfo{person}{Laurence A.~F. Park}.} \bibinfo{year}{2009}\natexlab{}.
\newblock \showarticletitle{Score Adjustment for Correction of Pooling Bias}.
  In \bibinfo{booktitle}{\emph{Proc. of SIGIR}}.
\newblock


\bibitem[\protect\citeauthoryear{Wolf, Debut, Sanh, Chaumond, Delangue, Moi,
  Cistac, Rault, Louf, Funtowicz, Davison, Shleifer, von Platen, Ma, Jernite,
  Plu, Xu, Le~Scao, Gugger, Drame, Lhoest, and Rush}{Wolf
  et~al\mbox{.}}{2020}]%
        {wolf2019huggingface}
\bibfield{author}{\bibinfo{person}{Thomas Wolf}, \bibinfo{person}{Lysandre
  Debut}, \bibinfo{person}{Victor Sanh}, \bibinfo{person}{Julien Chaumond},
  \bibinfo{person}{Clement Delangue}, \bibinfo{person}{Anthony Moi},
  \bibinfo{person}{Pierric Cistac}, \bibinfo{person}{Tim Rault},
  \bibinfo{person}{Remi Louf}, \bibinfo{person}{Morgan Funtowicz},
  \bibinfo{person}{Joe Davison}, \bibinfo{person}{Sam Shleifer},
  \bibinfo{person}{Patrick von Platen}, \bibinfo{person}{Clara Ma},
  \bibinfo{person}{Yacine Jernite}, \bibinfo{person}{Julien Plu},
  \bibinfo{person}{Canwen Xu}, \bibinfo{person}{Teven Le~Scao},
  \bibinfo{person}{Sylvain Gugger}, \bibinfo{person}{Mariama Drame},
  \bibinfo{person}{Quentin Lhoest}, {and} \bibinfo{person}{Alexander Rush}.}
  \bibinfo{year}{2020}\natexlab{}.
\newblock \showarticletitle{Transformers: State-of-the-Art Natural Language
  Processing}. In \bibinfo{booktitle}{\emph{Proc. EMNLP: System
  Demonstrations}}. \bibinfo{pages}{38--45}.
\newblock


\bibitem[\protect\citeauthoryear{Xiong, Dai, Callan, Liu, and Power}{Xiong
  et~al\mbox{.}}{2017}]%
        {Xiong2017}
\bibfield{author}{\bibinfo{person}{Chenyan Xiong}, \bibinfo{person}{Zhuyun
  Dai}, \bibinfo{person}{Jamie Callan}, \bibinfo{person}{Zhiyuan Liu}, {and}
  \bibinfo{person}{Russell Power}.} \bibinfo{year}{2017}\natexlab{}.
\newblock \showarticletitle{{End-to-End Neural Ad-hoc Ranking with Kernel
  Pooling}}. In \bibinfo{booktitle}{\emph{Proc. of SIGIR}}.
\newblock


\bibitem[\protect\citeauthoryear{Xiong, Xiong, Li, Tang, Liu, Bennett, Ahmed,
  and Overwijk}{Xiong et~al\mbox{.}}{2020}]%
        {xiong2020approximate}
\bibfield{author}{\bibinfo{person}{Lee Xiong}, \bibinfo{person}{Chenyan Xiong},
  \bibinfo{person}{Ye Li}, \bibinfo{person}{Kwok-Fung Tang},
  \bibinfo{person}{Jialin Liu}, \bibinfo{person}{Paul Bennett},
  \bibinfo{person}{Junaid Ahmed}, {and} \bibinfo{person}{Arnold Overwijk}.}
  \bibinfo{year}{2020}\natexlab{}.
\newblock \showarticletitle{Approximate Nearest Neighbor Negative Contrastive
  Learning for Dense Text Retrieval}.
\newblock \bibinfo{journal}{\emph{arXiv preprint arXiv:2007.00808}}
  (\bibinfo{year}{2020}).
\newblock


\bibitem[\protect\citeauthoryear{Zhan, Mao, Liu, Guo, Zhang, and Ma}{Zhan
  et~al\mbox{.}}{2021}]%
        {zhan2021jointly}
\bibfield{author}{\bibinfo{person}{Jingtao Zhan}, \bibinfo{person}{Jiaxin Mao},
  \bibinfo{person}{Yiqun Liu}, \bibinfo{person}{Jiafeng Guo},
  \bibinfo{person}{Min Zhang}, {and} \bibinfo{person}{Shaoping Ma}.}
  \bibinfo{year}{2021}\natexlab{}.
\newblock \showarticletitle{Jointly optimizing query encoder and product
  quantization to improve retrieval performance}. In
  \bibinfo{booktitle}{\emph{Proc. of CIKM}}.
\newblock


\bibitem[\protect\citeauthoryear{Zhou and Devlin}{Zhou and Devlin}{2021}]%
        {zhou2021multi}
\bibfield{author}{\bibinfo{person}{Giulio Zhou} {and} \bibinfo{person}{Jacob
  Devlin}.} \bibinfo{year}{2021}\natexlab{}.
\newblock \showarticletitle{Multi-Vector Attention Models for Deep Re-ranking}.
  In \bibinfo{booktitle}{\emph{Proceedings of the 2021 Conference on Empirical
  Methods in Natural Language Processing}}. \bibinfo{pages}{5452--5456}.
\newblock


\bibitem[\protect\citeauthoryear{Zobel}{Zobel}{1998}]%
        {zovel1998reliability}
\bibfield{author}{\bibinfo{person}{Justin Zobel}.}
  \bibinfo{year}{1998}\natexlab{}.
\newblock \showarticletitle{How Reliable Are the Results of Large-Scale
  Information Retrieval Experiments?}. In \bibinfo{booktitle}{\emph{Proc. of
  SIGIR}}.
\newblock


\end{thebibliography}
